\documentclass[reprint, amsmath,amssymb,aps,twocolumn, prb,superscriptaddress]{revtex4-1}
\usepackage{amsfonts,amssymb}
\usepackage[subnum]{cases}
\usepackage{mathrsfs}
\usepackage{amsmath}
\usepackage[none]{hyphenat}
\usepackage{graphicx}
\usepackage{hyperref}    
\usepackage{bm}          
\usepackage{natbib}
\usepackage{soul} 
\usepackage{color}
\usepackage{longtable}
\usepackage{ textcomp }
\usepackage{verbatim}
\usepackage{braket}
\usepackage{subfigure}

\DeclareTextFontCommand{\textroman}{\fontlibertine}

\makeatletter

\newcommand{\Rmnum}[1]{\expandafter\@slowromancap\romannumeral #1@}

\makeatother
\begin{document}
\title{Universal Symmetry-Protected Resonances in a Spinful Luttinger Liquid}
\author         {Yichen Hu, C. L. Kane}
\affiliation    {Department of Physics and Astronomy, University of Pennsylvania, Philadelphia, PA 19104}
\date{\today}

\begin{abstract}
We study the problem of resonant tunneling through a quantum dot in a spinful Luttinger liquid. For a range of repulsive interactions, we find that for symmetric barriers there exist resonances with a universal peak conductance $2g^* e^2/h$ that are controlled by a non-trivial intermediate fixed point. This fixed point is also a quantum critical point separating symmetry-protected topological phases. By tuning the system through resonance, all symmetry protected topological phases can be accessed. For a particular interaction strength with Luttinger parameters $g_\rho=1/3$ and $g_\sigma=1$, we show that the problem is equivalent to a two channel $SU(3)$ Kondo problem($SU(3)_2$ CFT). At the Toulouse limit, both problems can be mapped to a quantum Brownian motion model on a Kagome lattice, which in turn is related to the quantum Brownian motion on a honeycomb lattice and the three-channel $SU(2)$ Kondo problem($SU(2)_3$ CFT). Level-rank duality in the quantum Brownian motion model relating $SU(2)_k$ CFT to $SU(k)_2$ CFT is also explored. Utilizing the boundary conformal field theory, the on-resonance conductance of our resonant tunneling problem is calculated as well as the scaling dimension of the leading relevant operator. This allows us to compute the scaling behavior of the resonance line-shape as a function of temperature.

\end{abstract}

\maketitle

\section{INTRODUCTION}
Symmetry and topology are two foundational principles that shape our understanding of matter.    In the last decade, our understanding of their interplay has led to dramatic progress in our understanding of topological electronic phases.    A hallmark of this development is the topological insulator.  Insulating states with time reversal symmetry fall into two distinct topological phases that are separated by a topological quantum critical point\cite{HK,*QZ}.   For non-interacting systems described by the band theory, our understanding of such topological phenomena is highly developed, and there are many well understood examples of topological states protected by different types of symmetry, such as time-reversal symmetry\cite{KM,*KM2,*BZ}, particle-hole symmetry\cite{HK,*QZ}, spin rotation symmetry\cite{HK,*QZ}, as well as crystal symmetries\cite{F}.    A current frontier is to develop a similar understanding of symmetry protected topological phenomena in strongly interacting systems\cite{CGW,*LW,*CGLW}.    Since the general many-body problem is notoriously difficult, a promising approach is to consider the simplest version of a strongly interacting symmetry protected topological state:  one which occurs in a $0+1$ dimensional quantum impurity problem.   The archetypal quantum impurity problem is the Kondo problem\cite{KD}, along with its multichannel variants\cite{NB}. Previous works have explored the Kondo physics in closely related problems, such as resonant tunneling in non-Abelian quantum Hall states coupled to a quantum dot\cite{FBN,*FBN2,FFN,*FFN2}, fractional quantum Hall/normal-metal junctions in the strongly coupling regime\cite{SF} and resonant tunneling through a weak link in an interacting one dimensional electron gas - or a Luttinger liquid \cite{AY,MUW,KF2,FN,KF}.  

In this paper, we revisit the resonant tunneling problem in a Luttinger liquid.   This problem was studied extensively in the 1990's\cite{KF2,FN,KF}, where it was found that for spinless electrons with repulsive interactions (described by a Luttinger parameter $g$ with $g<1$) an arbitrarily weak barrier leads to an insulating behavior in the limit of zero temperature.   However, for $1/4<g<1$ it is possible, by tuning two parameters, to achieve a resonance with perfect conductance at zero temperature.   At small but finite temperature, the line shape of the resonance is described by a universal crossover scaling function that connects two renormalization group fixed points: the perfectly transmitting (small barrier) fixed point and the perfectly reflecting (large barrier) fixed point.   It was further observed that for symmetric barriers, a perfect resonance could be achieved by tuning only a \textit{single} parameter.   Here we observe that this is an example of symmetry protected topological critical point separating two topologically distinct symmetry protected insulating states.    In the presence of inversion symmetry a one dimensional insulator is characterized by a quantized polarization, which takes two values: $P = 0$ mod $e$ or $P=e/2$ mod $e$\cite{HPB}.    Likewise, for our resonant tunneling problem, we can define a polarization $N$ mod $e$ to distinguish different symmetry protected topological phases. $N$ is the number of charges transferred across the infinite barrier in the large-barrier limit. Without an inversion symmetric barrier, $N$ can take any continous value. With an inversion symmetric barrier, in the large barrier limit, $N=0$ or $e/2$ mod $e$, which characterizes two insulating phases. These insulating states are topologically distinct: one can not go smoothly from one phase to the other without going through a topological quantum critical point - the perfectly transmitting fixed point. For $1/4<g<1$, this critical point has only a single relevant operator in the presence of inversion symmetry.    For the special value $g=1/2$, this fixed point can also be identified with the non-Fermi liquid fixed point of the two-channel Kondo problem, described by a $SU(2)_2$ conformal field theory\cite{AFF, *AFF2}.

Armed with this insight we consider resonances in a spinful Luttinger liquid, which will lead us to a class of symmetry protected resonance fixed points that was not studied in detail in the early work.    A spinful Luttinger liquid is characterized by two Luttinger parameters $g_{\rho}$ and $g_{\sigma}$, with $SU(2)$ spin symmetry fixing $g_{\sigma} = 1$ \footnote{The value of Luttinger parameters $g_{\rho}$ and $g_{\sigma}$ is set to $2$ for noninteracting electrons in spinful Luttinger liquid in Ref. \onlinecite{KF2,KF}}.   As shown in Ref. \onlinecite{KF2,KF}, the system can achieve perfect resonance by tuning a single parameter for $1/2 < g_{\rho} < 1$. This resonance, which is controlled by the perfectly transmitting fixed point, corresponds to a transition between insulating phases characterized by a polarization $N_{pair}$. This polarization reflects whether or not a pair of electrons with opposite spins is transferred across the infinite barrier in the large-barrier limit. With inversion and time-reversal symmetry, the only possible values of $N_{pair}$ are $0$ or $e$ mod $2e$. When $g_{\rho}<1/2$, the perfectly transmitting fixed point becomes unstable even on resonance. In that case a new kind of insulating phase emerges characterized by $N_{pair}=\pm e/2$ mod $2e$. Even though, like the other two insulating phases, the new phase is charge insulating, with time-reversal symmetry, the spin degree of freedom in this phase is not completed locked due to the fact that an unpaired spin can be transferred across resulting in a finite conductance for spin. 
Transitions between these insulating states are governed by a quantum critical point that can not be described by a free Luttinger liquid fixed point.   Rather, it is an \textit{intermediate} fixed point\cite{YK,*Y}, which could only be described in certain perturbative limits.    Here we will show that like the spinless case there is a special value of $g_{\rho} = 1/3$ for which the nontrivial fixed point maps to a two-channel $SU(3)$ Kondo problem, described by a $SU(3)_2$ conformal field theory\cite{AOS,*AOS2}.    This analysis allows us to compute the nontrivial on-resonance conductance, as well as the scaling behavior of the width of the resonance as a function of temperature, which is determined by the dimension of the leading relevant operators at the nontrivial fixed point.

\begin{figure}[htbp]
\centering 
 \includegraphics[width=3.5in]{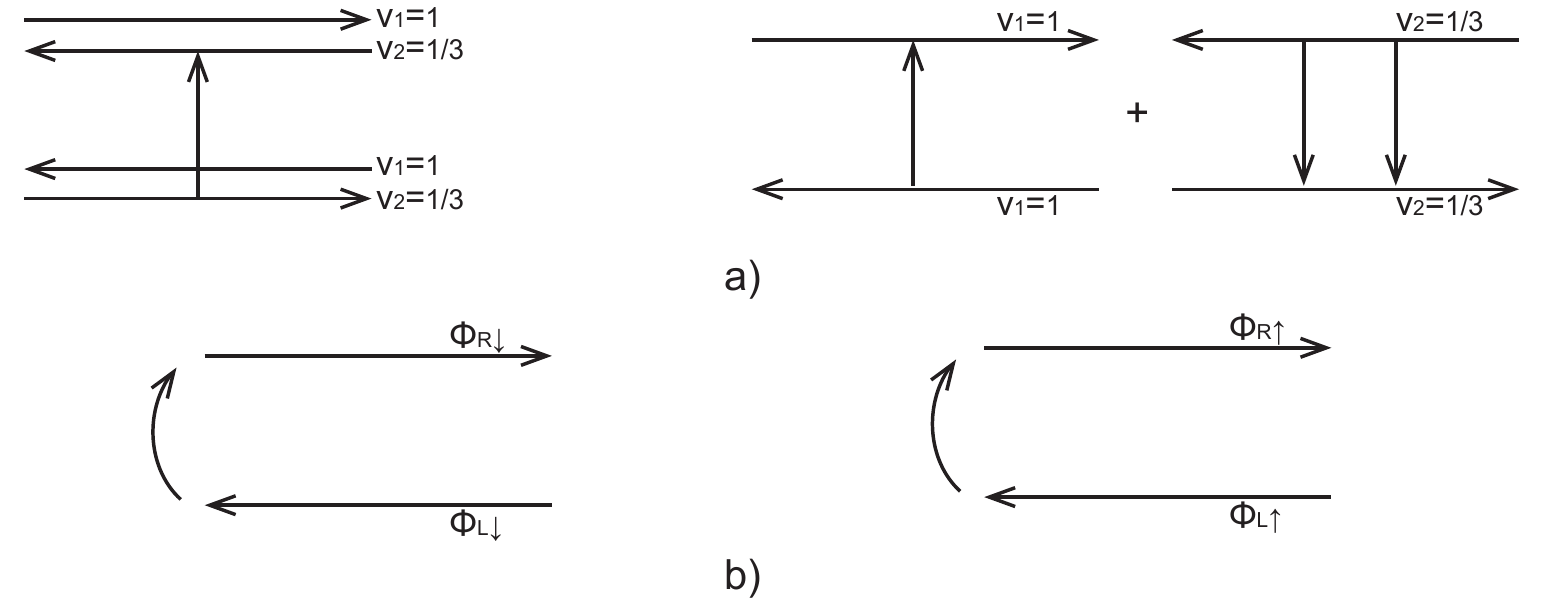}
\caption{a) Two $e=1/3$ quasiparticle tunneling processes in a $\nu=2/3$ fractional quantum Hall system. b) Two backscattering processes in a spinful Luttinger liquid.}
\end{figure}

We also note that the special point $g_{\rho}=1/3$ of the 1D spinful Luttinger liquid model is also of direct relevance to a corresponding resonant tunneling problem between edge states in the fractional quantum Hall effect, for which the Luttinger parameter is not an interaction dependent quantity.   Specifically, at filling $\nu=2/3$, disorder is predicted to lead to an edge state that has an \textit{upstream} neutral mode with an emergent $SU(2)$ symmetry\cite{KFP}. Further calculations\cite{MKGF} show that the backscattering terms of an electron in a Luttinger liquid can be identified as the tunneling terms of an $e=1/3$ quasiparticle in the fractional quantum Hall system (Fig. 1). In this case, we show that the problem of resonant tunneling through symmetric barriers is controlled by the $SU(3)_2$ fixed point.   

Insight into the relationships between the resonant tunneling problem and the Kondo problem is provided by mapping both problems to a quantum Brownian motion model\cite{CL,FZ,S,GHM}.   In the case of a spinful Luttinger liquid, the single impurity problem maps to a quantum Brownian motion on a two dimensional lattice.   We will argue that both the resonant tunneling problem and the two-channel $SU(3)$ Kondo problem are described by the quantum Brownian motion on a \textit{Kagome} lattice, when they are tuned to an appropriate Toulouse limit\cite{T}.   We will show that this, in turn is closely related to the quantum Brownian motion on a \textit{honeycomb} lattice, which was shown earlier to be related to the three-channel $SU(2)$ Kondo problem.   We will argue that this quantum Brownian motion picture provides a new insight into the level-rank duality that relates the $SU(3)_2$ and $SU(2)_3$ conformal field theories.

The paper is organized as follows. In Section II, we review symmetry protected topological phases for both the spinless and spinful Luttinger liquid. In Section III, we analyze our resonant tunneling problem which maps to a quantum Brownian motion model on a Kagome lattice at the Toulouse limit. From the quantum Brownian motion model, an intermediate fixed point is identified. Then we show how our resonant tunneling problem maps to a two-channel $SU(3)$ Kondo problem which comes handy for later analysis of the same fixed point. In Section IV, utilizing the boundary conformal field theory, we calculate the on-resonance conductance and by identifying the ``knob" controlling resonance we determine the critical exponent determining the scaling of the resonance line-shape with temperature.   In Section V, we show that the quantum Brownian motion on both the honeycomb lattice and the Kagome lattice flows to the same fixed point characterized by its mobility which manifests the so called level-rank duality.   We also point out some generalizations.

\section{Symmetry-protected topological phases in resonant tunneling problem}
Let us first take a look at the resonant tunneling problem in a spinless Luttinger liquid\cite{KF}. Introduced by Haldane\cite{HAL,*HAL2}, for spinless electrons, we can represent electrons in terms of two bosonic fields $\theta$ and $\varphi$ with the following commutation relation 
\begin{equation}
[\partial_x \theta(x),\varphi(x')]=i \pi \delta(x-x').
\end{equation}
The fermionic charge operators can then be bosonized as
\begin{equation}
\psi(x)\approx \sum_{m\hspace{1 mm} \text{odd}} e^{im(k_F x+\theta(x))}e^{i\varphi(x)},
\end{equation} where $k_F$ is the Fermi momentum and the effective Hamiltonian density may be written as
\begin{equation}
\mathcal{H}=\frac{v_F}{2\pi}[g(\partial_x\varphi)^2+\frac{1}{g}(\partial_x \theta)^2]
\end{equation} where $v_F$ is the sound velocity and $g$ is the Luttinger parameter characterizing strength of interaction, with $g=1$ corresponding to noninteracting fermions. The Euclidean action is $S_0=\int dx d\tau (i/\pi) \partial_\tau \varphi \partial_x \theta +\mathcal{H}$. Integrating our either $\theta$ or $\varphi$, we have two equivalent dual representations. 

At the small-barrier limit, integrating over $\varphi$ gives 
\begin{equation}
S_0=\int dx d\tau \frac{1}{2\pi g}[v(\partial_x\theta)^2+v^{-1}(\partial_{\tau} \theta)^2].
\end{equation}
This representation of the action is particularly convenient. Scattering of electrons from coupling to a potential $V(x)$ adds $\delta H=\int dx V(x) \psi^\dagger (x) \psi(x)$ to the Hamiltonian. Assuming $\theta(x)$ varies slowly on the scale of the potential and $V(x)$ is nonzero only near $x=0$, integrating out fluctuations in $\theta(x)$ away from zero, the effective action becomes
\begin{equation}
S_{0}=\frac{1}{\pi g}\sum_{i\omega_n} |\theta(\omega_n)|^2,
\end{equation}
plus an extra term corresponding to the effect of the potential:
\begin{equation}
-\int d\tau \frac{1}{2} \sum_{n=-\infty}^{\infty} v_n e^{i2n\theta(x=0,\tau)}
\end{equation} where $v_n=v_{-n}^*$ are Fourier components of $V(x)$ at momenta $2nk_F$ and $\omega_n$ is the Matsubara frequency. The extra term serves as the effective weak pinning potential for our resonant tunneling problem and we denote it as $V_{\text{eff}}[\theta(x=0)]$. To leading order in the backscattering, the RG flow equations are
\begin{equation}
\frac{dv_n}{d\ell}=(1-gn^2)v_n.
\end{equation}
Notice that for $1/4<g<1$, the only relevant perturbation is the backscattering term at $k=2k_F$: 
$$\text{Re}\, (v_1) \cos(2\theta)-\text{Im} \,(v_{1}) \sin(2\theta).$$ In general, the system achieves resonance by tuning the two  coefficients $\text{Re}(v_1)$ and $\text{Im}(v_1)$. With inversion symmetric barrier ($V(x)=V(-x)$), $v_1$ is a real number and therefore only one parameter needs to be tuned. 

At the opposite limit - the large-barrier limit, it is not hard to see that a convenient representation of the action should be in $\varphi$ variables since $\theta$s must be locked in the minimum of cosine potential and can not change continuously. Any perturbation away from this limit can be represented as hopping processes of electrons across the infinite barrier. We denote the strength of hopping as $t$. 

\begin{figure}[htbp]
\centering 
 \includegraphics[width=3in]{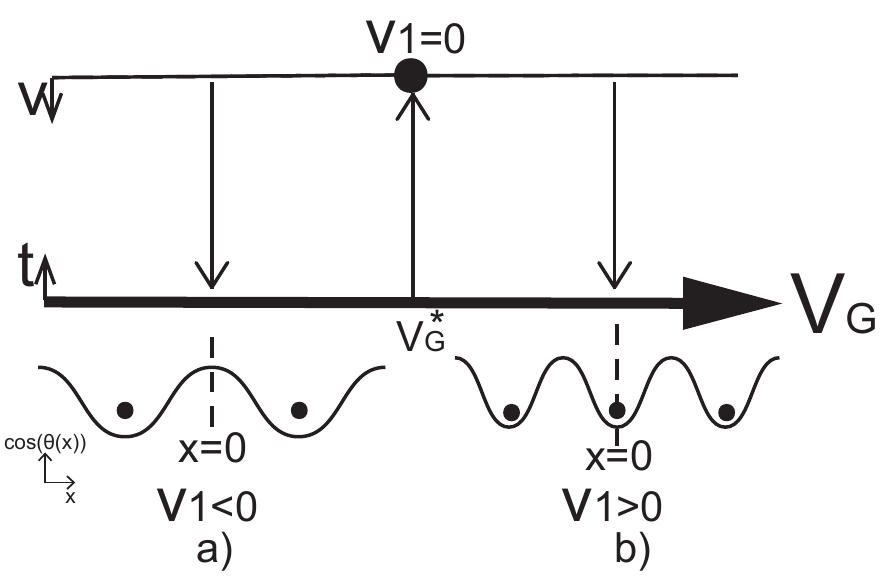}
\caption{Flow diagram for spinless resonant tunneling problem. The top(bottom) line represents small(large) barrier limits. Arrows represent RG flows and the solid dot represents the perfectly transmitting fixed point. At the large-barrier limit, two inversion symmetry protected insulating phases emerge shown in a) and b) represented in their cosine potential configuration. The dashed line indicates the center of inversion. $V_G$ is the gate voltage on the quantum dot one can tune to achieve resonance at $V_G^*$.}
\end{figure}

From Fig. 2, we see there are two inversion symmetry protected topological insulating phases for $1/4<g<1$ when the system flows into the large-barrier limit and they are separated by the perfect transmitting fixed point. If we choose the center of inversion as our origin, then for one insulating phase ($v_1>0$), with the infinite barrier, we must have $V_{\text{eff}}[\theta(x=0)]$ pinned in the potential minimum which is at $\theta(x=0)=\pi/2$. This corresponds to a polarization $N=e/2$ mod $e$. The other insulating phase ($v_1<0$) must have its minimum of the potential pinned at $\theta(x=0)=0$ and results in a polarization $N=0$ mod $e$. The two phases are topologically robust since the transition between them is only possible by tuning the system through resonance(tuning $v_1$ through 0). On resonance, we know the system is perfectly transmitting and electrons move freely. Therefore, our transition is analogous to the transition from a topological insulator to an ordinary insulator\cite{HK}. Both transitions go from one insulating phase to another insulating phase via a conducting state. Of course the conducting state for topological/ordinary insulator transition refers to the familiar band gap closure.

The resonant problem in a spinless Luttinger liquid provides the simplest example of a $0+1 d$ system with symmetry protected topological phases. The paradigm here is to recognize the perfect resonance fixed point as the symmetry-protected quantum critical point separating symmetry-protected topological phases.    

Adopting this new interpretation, let us now move on to the resonant problem in a spinful Luttinger liquid. For electrons with spin, we have two Luttinger parameters, the dimensionless conductance $g_{\rho}$ and the dimensionless ``spin conductance" $g_{\sigma}$ which describes the spin-current response to a magnetic field. 
For each spin $\mu=\uparrow, \downarrow$, there are two bosonic fields $(\theta_{\mu},\varphi_{\mu})$. It is convenient to separate them into charge and spin degrees of freedom:
\begin{equation}
\theta_{\rho}=\theta_{\uparrow}+\theta_{\downarrow},
\theta_{\sigma}=\theta_{\uparrow}-\theta_{\downarrow}.
\end{equation} In the small-barrier limit, there are two competing perturbation terms in the action which are most relevant for $g_{\rho}<1$ and $g_{\sigma}=1$\cite{KF}
\begin{equation}
-\int d\tau v_e \cos(\theta_\rho) \cos(\theta_\sigma)
\end{equation} and
\begin{equation}
-\int d\tau v_1\cos(2\theta_\rho)+v_2 \sin (2\theta_\rho),
\end{equation}
where $v_e$ is the process that backscatters an electron and $v_\rho=v_1+iv_2$ is the process that backscatters an up-spin and also a down-spin electron. These two perturbation terms combined is the effective weak pinning potential $V_{\text{eff}}(\theta_{\rho}(x=0), \theta_{\sigma}(x=0))$ of the spinful resonant tunneling problem and their flow equations are given as 
\begin{eqnarray}
\frac{d v_e}{d \ell}=(1-\frac{g_\rho}{2}-\frac{g_\sigma}{2})v_e\\
\frac{d v_\rho}{d \ell}=(1-2g_\rho)v_\rho.
\end{eqnarray}
Notice that the $v_{\rho}$ process is relevant only for $g_{\rho}<1/2$. There is another process $v_\sigma$ which corresponds to backscattering of an up spin and a down spin electron incidenting from opposite directions(net charge momentum unchanged). If this process is relevant, it could pin $\theta_\sigma$ to the minimum of potential. However, in the range of Luttinger parameters of our discussion, this process will always be irrelevant.\cite{KF}
\begin{figure}[htbp]
\centering 
\includegraphics[width=3in]{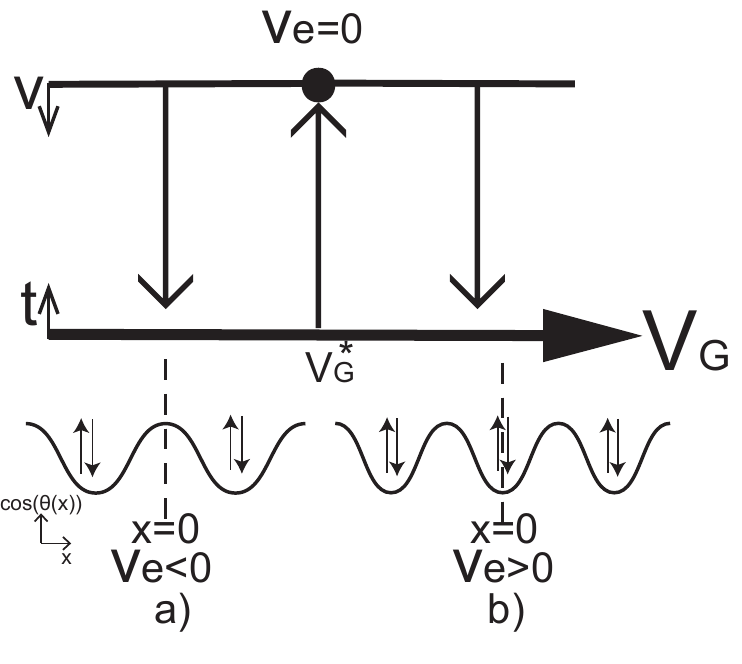}

\caption{Flow diagram for the spinful case with $1/2<g_{\rho}<1$ and $g_\sigma=1$. The top(bottom) line represents small(large) barrier limits. Arrows represent RG flows and the solid dot represents the perfectly transmitting fixed point. At the large-barrier limit, two inversion and time-reversal symmetry protected insulating phases emerge shown in a) and b) represented in their cosine potential configuration. The dashed line indicates the center of inversion. $V_G$ is the gate voltage on the quantum dot one can tune to achieve resonance at $V_G^*$.}
\end{figure}

Now, when $1/2<g_\rho<1 $ and $g_\sigma=1$, the $v_\rho$ process will be irrelevant. Two topologically distinct insulating phases separated by a perfectly transmitting fixed point again emerge as shown in Fig. 3. However, this time, they are protected by both inversion symmetry and time-reversal symmetry with the potential minimum pinned at $(\theta_\rho(x=0)=\pi, \theta_\sigma(x=0)=0)$ for $v_e>0$ and $(\theta_\rho(x=0)=0, \theta_\sigma(x=0)=0)$ for $v_e<0$ as the system flows into the large-barrier limit. They are characterized by $N_{pair}=e$ or $0$ mod $2e$ respectively. The transition between these two symmetry protected topological phases are achieved by tuning the system to resonance(tuning $v_e$ through $0$).

\begin{figure}[htbp]
\centering 
\includegraphics[width=3.5in]{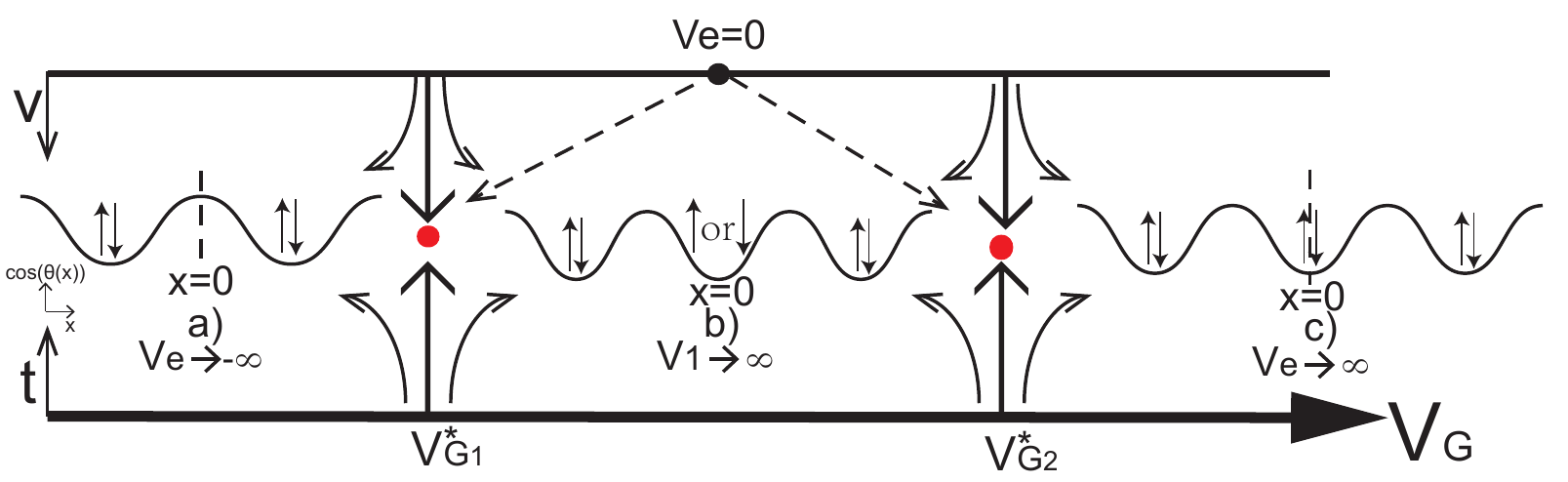}
 
\caption{For the spinful case with $g_{\rho}<1/2$ and $g_\sigma=1$, the perfectly transmitting fixed point(the black solid dot) becomes unstable and opens up to the intermediate fixed point(the red solid dots) as indicated by the dashed arrows. Under RG flows(indicated by arrows), three inversion and time-reversal symmetry protected phases emerge shown in a), b) and c) represented in their cosine potential configuration. $V_G$ is the gate voltage on the quantum dot one can tune to achieve resonance at $V_{G_{1,2}}^*$.}
\end{figure}

A more interesting intermediate fixed point can be found if we make interactions in charge sector more repulsive ($g_\rho<1/2$) while keep spin symmetry ($g_\sigma=1$).  The $v_\rho=v_1$ (due to inversion symmetry, the $v_2$ term is eliminated) process is now relevant, and we have to take it into account alongside the $v_e$ process. There are two different situations depending on the sign of $v_1$. When $v_1>0$,  the minimum of the potential $V_{\text{eff}}$ is pinned to $\theta_\rho(x=0)=\pi/2$. The $v_e$ process is thus eliminated and the $v_1$ process dominates and grows to infinity under RG flows. The previous perfectly transmitting fixed point becomes unstable in this occasion and flows into an intermediate fixed point. A new symmetry protected topological phase emerges as shown in Fig. 4. Note that at this new phase we are free to change $\theta_\sigma$ since the $v_{\sigma}$ process is still irrelevant. Thus, it is a charge insulating phase with a finite spin conductance characterized by $N_{pair}=\pm e/2$ mod $2e$. On the other hand, when $v_1<0$, the minimum of the potential is pinned to either $\theta_\rho(x=0)=0$ or $\pi$. In this case, since the $v_e$ process is also present and its magnitude grows to infinity under RG flows, ($\theta_{\rho}(x=0)$, $\theta_{\sigma}(x=0)$) will be locked to either $(0,0)$ or $(\pi,0)$ depending on the sign of $v_e$ as before. This gives us two charge and spin insulating phases. All three symmetry protected topological phases can be accessed by adjusting a single parameter - the ratio $v_e/v_1$. A previous study of this intermediate fixed point can be found in Ref. \onlinecite{KF2, KF}. It was shown that this fixed point becomes perturbatively accessible from the perfectly transmitting fixed point with an $\epsilon$-expansion near critical values of Luttinger parameters $g_{\rho}^*=1/2$ and $g_{\sigma}^*=3/2$ at the small-barrier limit. This is due to the fact that the perfectly transmitting fixed point becomes unstable at the aforementioned values of Luttinger parameters. Unfortunately, for our $SU(2)$ symmetric case with $g_{\sigma}=1$, this intermediate fixed point is not perturbatively accessible using the $\epsilon$-expansion method. However, for $g_\rho=1/3$, an exact description can be obtained by using boundary conformal field theory as shown in Section \Rmnum{4}.

\section{Resonant tunneling problem and related quantum impurity problems}
In this section, we will further develop our understanding of the resonant tunneling problem in spinful Luttinger liquid and explore the connection between our resonant tunneling problem and other quantum impurity problems. First, the simpler spinless resonant tunneling problem\cite{KF} is reviewed to pave the way for understanding the more complicated spinful case. 
Then,  we perform renormalization group calculations for our spinful resonant tunneling problem. At the Toulouse limit, our resonant tunneling problem is nothing but a quantum Brownian motion model on a Kagome lattice. At both small (small $v$) and large (small $t$) barrier limits of the quantum Brownian motion model, the system flows to an intermediate fixed point. To obtain an exact description of this fixed point, we map our resonant tunneling problem to a two-channel Kondo problem with $SU(3)$ impurity spin\cite{AOS,*AOS2}.

\subsection{SPINLESS RESONANT TUNNELING PROBLEM}
\begin{figure}[htbp]
\centering 
 \includegraphics[width=3.5in]{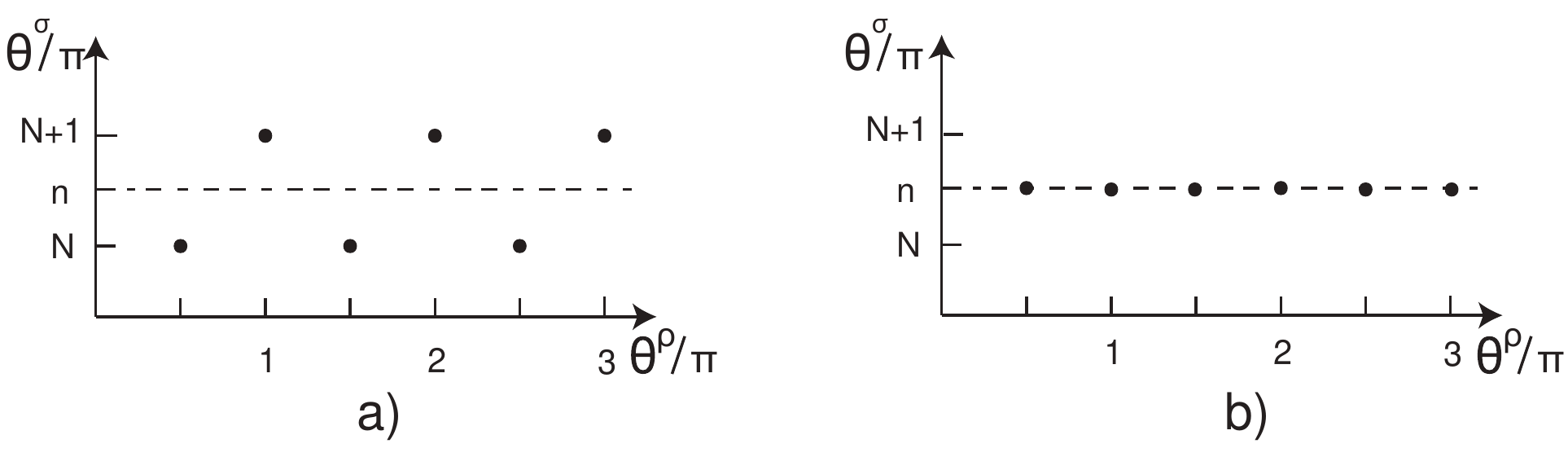}
\caption{Positions of the minima of the action in the $\theta^{\rho}$-$\theta^{\sigma}$ plane. a) $K=1$, b) $K=0$.}
\end{figure}

Again we start with the spinless resonant tunneling problem. Taking the large-barrier limit, if the capacity on the quantum dot is small, a large charging energy fixes the number of charge on the dot and transmissions through the dot are suppressed. By tuning the gate voltage, the chemical potential on the dot can be adjusted and resonant tunneling can be achieved. A theoretical model takes a double-barrier structure, it is a wire with two $\delta$-functions on it separated by a quantum dot with size $d$. On the dot, a gate voltage $V_G$ is assigned. We denote $\theta^{i=1,2}/\pi$ as the number of electrons tunneling through the corresponding barrier. We also define $\theta^{\rho}/\pi=(\theta^{1}+\theta^{2})/\pi$ as the number of electrons transferred across two barriers and $\theta^{\sigma}/\pi=(\theta^{2}-\theta^{1})/\pi$ as the number of electrons on the dot. Then,the action has deep minima when $\theta^{\sigma}/\pi$ is an integer (see Fig. 5a). Since for infinite large barriers $\theta$ fields are pinned at minima, it is more convenient to use the $\varphi$ representation($(\varphi^{\rho},\varphi^\sigma)$ are dual bosonic fields of $(\theta^\rho,\theta^\sigma)$ following from the standard bosonization terminology). In this case, the partition function describes instantons connecting these degenerate minima. The hopping processes of instantons correspond physically to electrons hopping on or off the quantum dot. The partition function can be analyzed in the Coulomb-gas representation in powers of the tunneling amplitude $t$,
\begin{equation}
\begin{aligned}
Z&=\sum_n \sum_{\{q=\pm 1\}} \int \frac{d^{2n} \tau}{\tau_c}  \langle (t\ket{1}\bra{0} e^{-i(q\frac{1}{\sqrt{g}} \varphi^{\rho}+\frac{\sqrt{K}}{\sqrt{g}} \varphi^{\sigma})}\\
&+\text{h.c.})^{2n}\rangle.
\end{aligned} 
\end{equation} where $\ket{1},\ket{0}$ are quantum states on the dot labeling the number of electrons on the dot. $K$ is the renormalization constant and is initially set to be $1$. Its value flows under renormalization group.

Integrating out bosonic fields $\varphi$ mediates a logarithmic interaction between ``charges" in the Coulomb-gas representation. The ``charges" correspond to physical hopping processes and we have two kind of ``charges" in our problem: hopping electrons on and off the dot. After integration, the partition function is in the following form:
\begin{equation}
\begin{aligned}
Z&=\sum_n \sum_{\{q_i=\pm 1\}} t^{2n} \int \frac{d^{2n}}{\tau_c} e^{-\sum_{i<j} V_{ij}}, \\
V_{ij}&=\frac{2}{g} (q_i q_j +K r_i r_j) \ln{\frac{(\tau_i-\tau_j)}{\tau_c}},
\end{aligned}
\end{equation} where $q_i=\theta^{\rho}/\pi=\pm 1$ denotes the charge transferred to the right in a hopping event and $r_i=\theta^{\sigma}/\pi=\pm 1$ denotes the change in charge on the dot. Due to the discreteness of charge on the dot which can only change by 1, $r_i$ must alternate whereas $q_i$ can have any ordering.
 
Tuning into resonance, the system renormalizes according to the RG flow equations\cite{KF}
\begin{equation}
\begin{aligned}
&\frac{dK}{dl}=-8\tau_c^2 t^2 K,\\
& \frac{dt}{dl}=t[1-\frac{(1+K)}{4g}].
\end{aligned}
\end{equation}During the process, electrons on the dot can virtually tunnel back to the leads reducing the average charge on the dot $\theta^{\sigma}$. On resonance, the hopping directions along $\theta^{\sigma}$ collapse($K=0$) which renders $\theta^{\sigma}/\pi=n$ into precisely a half integer(Fig. 5b). This reduction of dimensionality from two to one at the Toulouse limit greatly simplifies the problem. A similar simplification will arise in the more complicated spinful problem discussed below. 

It is also worth mentioning that at $g=1/2$, the spinless resonant tunneling problem can be mapped to a two-channel Kondo problem with $SU(2)$ impurity spin\cite{KG,EK}.

\subsection{SPINFUL RESONANT TUNNELING PROBLEM}
\begin{figure}[!htb]
\centering
\mbox{\subfigure{\includegraphics[width=1.7in]{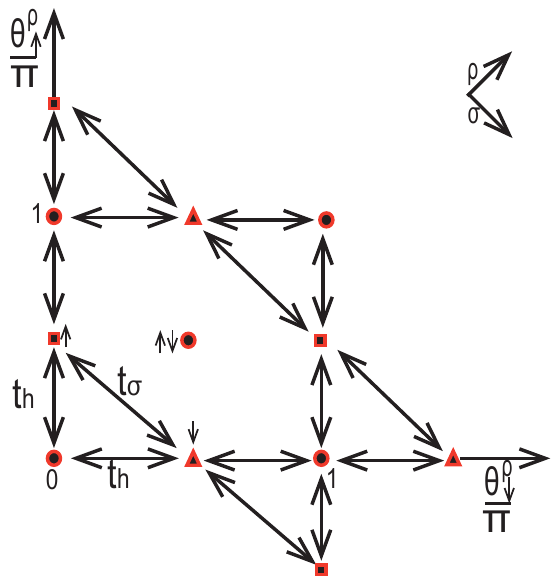}}
\subfigure{\includegraphics[width=1.7in]{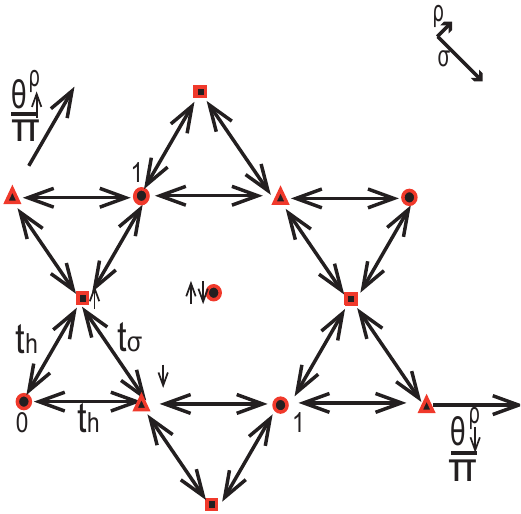}} }
\caption{Resonant tunneling problem in $\theta^\rho_\uparrow$-$\theta^\rho_\downarrow$ plane at the large-barrier limit. Red spots are minima of the periodic potential. a) $g_\rho=g_\sigma=1$. b) $g_\rho=1/3$ and $g_\sigma=1$. ($\bigcirc=\ket{0}$, $\square=\ket{\uparrow}$, $\triangle=\ket{\downarrow}$)} \label{fig12}
\end{figure}

Examining the spinful case, at the large-barrier limit, for each barrier, there are bosonic fields $(\theta,\varphi)$ as defined in Eq. (8). We can reorganize $(\theta,\varphi)$ fields into the following
\begin{equation}
\begin{aligned}
&\theta^{\rho}_{\rho,\sigma}=\frac{1}{\sqrt{2}}(\theta^1_{\rho,\sigma}+\theta^2_{\rho,\sigma}),
\theta^{\sigma}_{\rho,\sigma}=\frac{1}{\sqrt{2}}(\theta^1_{\rho,\sigma}-\theta^2_{\rho,\sigma}),\\
&\varphi^{\rho}_{\rho,\sigma}=\frac{1}{\sqrt{2}}(\varphi^1_{\rho,\sigma}+\varphi^2_{\rho,\sigma}),
\varphi^{\sigma}_{\rho,\sigma}=\frac{1}{\sqrt{2}}(\varphi^1_{\rho,\sigma}-\varphi^2_{\rho,\sigma}).
\end{aligned}
\end{equation} The superscript $\rho$ and $\sigma$ denote physical quantities transferred across two barriers or changed in the dot respectively. The subscript $\rho$ and $\sigma$ denote charge or spin respectively. Now, the action 
\begin{equation}
S=S_0+\int d\tau V_{\text{eff}}((\theta^{\rho}_{\uparrow}(\tau),\theta^{\rho}_{\downarrow}(\tau),\theta^{\sigma}_{\uparrow}(\tau),\theta^{\sigma}_{\downarrow}(\tau)))
\end{equation}($V_{\text{eff}}$ is a periodic potential possesses lattice symmetry shown in Fig. 6) will have deep minima whenever $\theta^{\rho}_{\uparrow}/\pi$(the number of electrons with up spin transferred over two barriers) or $\theta^{\rho}_{\downarrow}/\pi$  (the number of electrons with down spin transferred over two barriers) is an integer(Fig. 6). We can adopt the same Coulomb-gas representation of the partition function to describe our resonant tunneling problem. In our case there are three tunneling processes. The processes in which a spin up or down electron hops on or off the quantum dot has a tunneling amplitude $t_{h_\uparrow}$ or $t_{h_\downarrow}$. The other processes in which both the spin of an electron on the lead and that of an electron on the quantum dot are flipped has a tunneling amplitude $t_{\sigma}$. When $g_{\rho}=1/3$ and $g_{\sigma}=1$, the three tunneling processes have the same tunneling amplitude $t_{\sigma}=t_{h_{\uparrow,\downarrow}}=t$(Fig. 6).

Expanding the partition function in powers of $t_{\sigma}$ and $t_{h_{\uparrow,\downarrow}}$, we arrive at
\begin{equation}
\begin{aligned}
Z&=\sum_n \sum_{a=\pm} \int \frac{d^n\tau_j}{\tau_c} \langle (t_{\sigma} \ket{\downarrow} \bra{\uparrow } e^{-i[a\frac{1}{\sqrt{g_{\sigma}}} \varphi_{\sigma}^{\rho}+\frac{K_{\sigma}}{\sqrt{g_\sigma}} \varphi^{\sigma}_{\sigma}]}\\
&+t_{h_\uparrow} \ket{\uparrow}  \bra{0}  e^{-i[a(\frac{1}{2\sqrt{g_{\rho}}} \varphi_{\rho}^{\rho}+\frac{1}{2\sqrt{g_{\sigma}}} \varphi_{\sigma}^{\rho})+(\frac{K_{\rho}}{2\sqrt{ g_{\rho}}} \varphi^{\sigma}_{\rho}+\frac{K_{\sigma}}{2\sqrt{ g_{\sigma}}} \varphi^{\sigma}_{\sigma})]}\\
&+t_{h_{\downarrow}} \ket{\downarrow} \bra{0} e^{-i[a(\frac{1}{2\sqrt{g_{\rho}}} \varphi_{\rho}^{\rho}-\frac{1}{2\sqrt{g_{\sigma}}} \varphi_{\sigma}^{\rho})+(\frac{K_{\rho}}{2\sqrt{ g_{\rho}}} \varphi^{\sigma}_{\rho}-\frac{K_{\sigma}}{2\sqrt{ g_{\sigma}}} \varphi^{\sigma}_{\sigma})]}\\
&+\text{h.c.})^n\rangle\\
&=\sum_n \sum_{a=\pm} \int \frac{d^n\tau_j}{\tau_c} \langle
(\sum_{k=\sigma,h_{\uparrow},h_{\downarrow}} t_k \delta^{k+} e^{-i (a\vec{H_{k+}} \cdot \vec{\varphi^{\rho}}+\vec{H_{k+}} \textbf{K} \vec{\varphi^{\sigma}})}\\
&+\text{h.c.})^n \rangle
\end{aligned}
\end{equation}
where $\ket{l}\bra{m}_{l\neq m, l,m=\uparrow,\downarrow,0}$ have been relabeled as $\delta^{k\pm}_{k=\sigma, h_{\uparrow}, h_{\downarrow}}$ and exponents are shortened as dot products of vectors $\vec{H_{k\pm}}_{k=\sigma, h_{\uparrow}, h_{\downarrow}}$ 
\begin{eqnarray}
\vec{H_{\sigma_\pm}}&=\pm(0,\frac{1}{\sqrt{g_{\sigma}}})\\
\vec{H_{h_{\uparrow \pm}}}&=\pm(\frac{1}{2\sqrt{g_{\rho}}},\frac{1}{2\sqrt{g_{\sigma}}})\\
\vec{H_{h_{\downarrow \pm}}}&=\pm(\frac{1}{2\sqrt{g_{\rho}}},-\frac{1}{2\sqrt{g_{\sigma}}}),
\end{eqnarray}
$\vec{\varphi^j}_{(j=\rho,\sigma)}=(\varphi_{\rho}^j,\varphi_{\sigma}^j)$ and $\textbf{K}=
\begin{pmatrix}
K_{\rho} & 0\\
0 & K_{\sigma}
\end{pmatrix}$.

For $g_{\rho}=1/3$ and $g_{\sigma}=1$, we have $K_{\rho,\sigma}=K$ which is initially set to $1$ and $t_{\sigma,h_{\uparrow},h_\downarrow}=t$. Integrating out bosonic $\varphi$ fields mediates logarithmic interactions between ``charges" in the following form:
\begin{equation}
\begin{aligned}
Z&=\sum_n \frac{1}{n!} t^n \sum_{\{a_i=\pm 1\}} \int \frac{d^n\tau_i}{\tau_c} e^{-\sum_{i<j} V_{ij}},\\
V_{ij}&=(\vec{H_{i}} \textbf{K}^2 \vec{H_{j}}+a_i a_j \vec{H_{i}}\cdot\vec{H_{j}}) \ln{\frac{\tau_i-\tau_j}{\tau_c}}.
\end{aligned}
\end{equation}

Let us explain this Coulomb-gas model further. Because of the extra degree of freedom from spin, instantons now move in a four-dimensional space coordinated by $(\theta^{\rho}_{\rho},\theta^{\rho}_{\sigma},\theta^{\sigma}_{\rho}, \theta^{\sigma}_{\sigma})$ with discrete values. ``Charges" here are again different physical processes. Since there are six physical hopping processes: hopping on or off either an up or down electron to the dot and flipping the spin on the dot, relations among all possible processes for a single time step constitute a triangle(see Fig. 6). ``Charges" are now vectors of the triangle instead of scalars and their physical relevance are encoded in their length depending on Luttinger parameters. Three ``charges" are given in Eq. (19)-(21) in $(\theta^{\sigma}_{\rho},\theta^{\sigma}_{\sigma})$ coordinates characterizing the change of both spin and charge on the quantum dot like $r_i$ in the spinless case. At each time $\tau_j$, it has to alternate among all three possible states on the quantum dot. The other three ``charges" are $a_i \vec{H_i}$ in $(\theta^{\rho}_{\rho},\theta^{\rho}_{\sigma})$ coordinates analogous to $q_i$, characterizing both spin and charge transferred across two barriers with no restriction of alternation.

To put it more visually, imagine that we have two kinds of hopping on the four-dimensional lattice space, those perpendicular to hyper-surfaces with $(\theta_{\rho}^{\rho},\theta_{\sigma}^{\rho})$ coordinates held fixed and those parallel to hyper-surfaces with $(\theta_{\rho}^{\sigma},\theta_{\sigma}^{\sigma})$ coordinates held fixed. For the former case, instantons hop along $\vec{H_i}$s in the two dimensional sublattice with $(\theta_{\rho}^{\sigma},\theta_{\sigma}^{\sigma})$ and for the latter, instantons hop along $a_i \vec{H_i}$s in the two dimensional sublattice with $(\theta_{\rho}^{\rho},\theta_{\sigma}^{\rho})$. Here, $a_i=\pm 1$ since at each lattice site there are two corresponding vectors with opposite directions and instantons are free to choose one. Conservation of spin and charge in our resonant tunneling problem poses two constraints
\begin{eqnarray}
&\sum_{i=1}^2 \theta^i_{\rho}+\iota_{\rho}=\text{const}\\
&\sum_{i=1}^2 \theta^i_{\sigma}+\iota_{\sigma}=\text{const},
\end{eqnarray} where $\iota_{\rho, \sigma}$ are the charge and spin on the quantum dot. Thus, different values of $(\iota_{\rho},\iota_{\sigma})$ label hyper-surfaces in which tunneling processes take place. However, since spin or charge on the dot has to alternate among the three possible occupation states ($\ket{\uparrow}, \ket{\downarrow}, \ket{0}$), following the same reasoning as the spinless case, directions for tunneling between different hyper-surfaces along $(\theta^{\sigma}_{\rho},\theta^{\sigma}_{\sigma})$ will get renormalized and eventually leads to the four- dimensional lattice collapsing into the two-dimensional lattice shown in Fig. 6 with a change of basis to $(\theta^{\rho}_{\downarrow},\theta^{\rho}_{\uparrow})$.

\begin{figure}[htbp]
\centering 
 \includegraphics[width=3in]{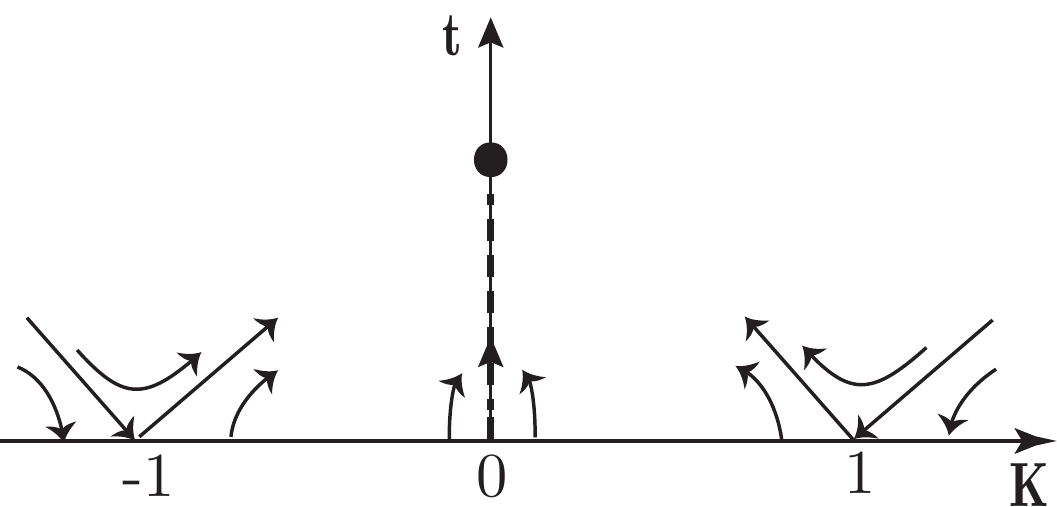}
\caption{Flow diagram for resonant tunneling problem in a spinful Luttinger liquid. The dashed line is the Toulouse limit($K=0$). The circle denotes the fixed point.}
\end{figure}

A detailed RG calculation in Appendix A gives the following  flow equations with the corresponding flow diagram Fig. 7. The Toulouse limit is along the line with $K=0$.
\begin{eqnarray}
&\frac{dt}{d\ell}=(1-\frac{1}{2}(K^2+1))t\\
&\frac{dK}{d\ell}=-6\tau_c^2 t^2K
\end{eqnarray}

This Toulouse limit($K=0$) of our resonant tunneling problem, where instantons are confined in a two dimensional sublattice with coordinates $(\theta_{\rho}^{\rho},\theta_{\sigma}^{\rho})$ and directions along the other two bosonic fields decouple, is identical to a quantum Brownian motion model on a \textit{Kagome} lattice.
More rigorously, at the Toulouse limit, the action for our resonant tunneling problem is 
\begin{equation}
\begin{aligned}
&S_{\text{RT}}=t\int \frac{d\tau}{\tau_c}\sum_{a=\pm 1}\sum_{i=\sigma,h_{\uparrow},h_{\downarrow}}[\delta^{i+} e^{-i a \vec{H_{i+}} \cdot \vec{\varphi^{\rho}}}+\text{h.c.}].
\end{aligned}
\end{equation}If we do the following mapping
\begin{eqnarray}
& t \leftrightarrow t_{\mathcal{\vec{R}}}\\
& {\delta^{i\pm}}_{i=\sigma,h_{\uparrow},h_{\downarrow}} \leftrightarrow \tau^{i\pm}_{i=\bigcirc,\square,\triangle}  \\
& \vec{\varphi^{\rho}}\leftrightarrow 2 \pi \vec{k}\\
& a\vec{H_{i\pm}} \leftrightarrow \vec{a\mathcal{R}_{i\pm}},
\end{eqnarray}
then this is precisely the action of a quantum Brownian motion model tunneling on a Kagome lattice in the large-barrier limit
\begin{equation}
S=t_{\mathcal{\vec{R}}}\int \frac{d\tau}{\tau_c} \sum_{a=\pm 1} \sum_{i=\bigcirc,\square,\triangle} [\tau^{i+} e^{i  \vec{a\mathcal{R}_{i\pm}} \cdot 2\pi\vec{k}(\tau)}+\text{h.c.}].
\end{equation} 
where $t_{\mathcal{\vec{R}}}$ is the amplitude of hopping between minima connected by a lattice vector $a\vec{\mathcal{R}_{i\pm}}$, $\vec{k}(\tau)$ is the position of particle in the momentum space with a potential possessing the symmetry of the reciprocal lattice and $\bigcirc,\square, \triangle$ describe lattice sites shown in Fig. 6.

The quantum Brownian motion model was originally proposed as a theoretical model for heavy charged particle in a metal\cite{CL}. Although the applicability of this model to its original proposed problem is questioned\cite{I,*ZVZ}, the model is later shown to be relevant to quantum impurity problems. It describes a Brownian particle moving in a lattice with a periodic potential. The coupling of the potential to the particle generates a frictional force which acts as dissipative bath.

There are two perturbatively accessible limits to analyze the effect of the periodic potential in a quantum Brownian motion model. In the small $v$ limit for which the barrier is small, the action is
\begin{equation}
S=S_0[\vec{l}(\tau)]-\int \frac{d\tau}{\tau_c}\sum_{\vec{G}} v_{\vec{G}} e^{i 2 \pi \vec{G} \cdot \vec{l}}
\end{equation} where $S_0$ is the dissipative kinetic energy and the latter integral represents the energy of the periodic potential. In the integrand the periodic potential amplitude at the particle trajectory $\vec{l}(\tau)$ is written in sums of Fourier components $v_{\vec{G}}$($\vec{G}$ is the reciprocal lattice vector). 

Under RG calculations in the leading order, the flow equation depends on the length of the reciprocal lattice vector 
\begin{equation}
\frac{dv_{\vec{G}}}{d\ell}=(1-|\vec{G}|)v_{\vec{G}}.
\end{equation}
Similarly, for the small t (large barrier) limit, the flow equation depends on the length of the lattice vector
\begin{equation}
\frac{dt_{\mathcal{\vec{R}}}}{d\ell}=(1-|\vec{\mathcal{R}}|)t_{\mathcal{\vec{R}}}
\end{equation}
We know that for a Kagome lattice the product of the shortest reciprocal lattice vector $|\vec{G_0}|$ and the shortest lattice vector $|\vec{R_0}|$ is $|\vec{G_0}| |\vec{R_0}|=1/ \sqrt{3}$. Then it follows that for $1/3<|\vec{G_0}|^2<1$, both small and large-barrier limits are \textit{unstable} and there must be a stable intermediate fixed point in between. The intermediate fixed point is characterized by the mobility $\mu$ of the Brownian particle under the external frictional force where $\mu=1$ at $v=0$ and $\mu=0$ at $t=0$. Thus $0<\mu^*<1$ for our intermediate fixed point. In general, $\mu^*$ depends on $|\vec{G_0}|^2$ and is hard to calculate. 

This reminds us of a previous work by Yi and Kane\cite{YK,*Y}. In it, they were able to map a quantum Brownian motion model on a $N-1$ dimension honeycomb lattice to the Toulouse limit of an $N$-channel Kondo problem with $SU(2)$ impurity spin. Like our model, for $4/9<|\vec{G_0}|^2<1$, the system flows into an intermediate fixed point characterized by the mobility $\mu^*$. When $|\vec{R_0}|^2=|\vec{G_0}|^2=2/3$, an exact description of this fixed point is possible from boundary conformal field theory since it is the same intermediate fixed point of the aforementioned three-channel $SU(2)$ Kondo problem. In order to find an exact description of the intermediate fixed point of our problem, we should again walk down this route of mapping to the multichannel Kondo problem. 

\begin{figure}[htbp]
\centering 
 \includegraphics[width=3in]{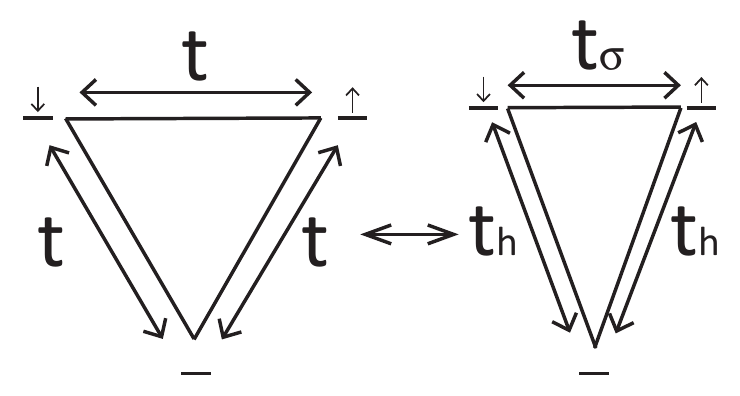}
\caption{Representative triangles of tunneling processes for both the special $g_\rho=1/3$ and $g_\sigma=1$ case (left) and the more general case (right). }
\end{figure}

Before we start our study of the multichannel Kondo problem, we should note that there is a more general situation for our resonant tunneling problem. With inversion and time-reversal symmetry, it is not guaranteed that the three tunneling processes have the same amplitude, however, although it does require $t_{h_\uparrow}=t_{h_\downarrow}=t_h$(Fig. 8). As a result, $K_{\rho}\neq K_{\sigma}$, RG flow equations are also modified as:
\begin{eqnarray}
&\frac{dt_{\sigma}}{d\ell}=(1-\frac{1}{2}(K_{\sigma}^2+1))t_{\sigma},\\
&\frac{dt_{h}}{d\ell}=(1-\frac{1}{2}(\frac{3}{4}K_{\rho}^2+\frac{1}{4}K_{\sigma}^2+1))t_{h},\\
&\frac{dK_{\sigma}}{d\ell}=-2((t_{h})^2+2(t_{\sigma})^2)\tau_c^2 K_{\sigma},\\
&\frac{dK_{\rho}}{d\ell}=-6(t_h)^2 \tau_c^2 K_{\rho}.
\end{eqnarray}

\subsection{CONNECTIONS TO MULTICHANNEL KONDO PROBLEM}
In this subsection, we will establish the equivalence between resonant tunneling problems in a Luttinger liquid and the multichannel Kondo problem. This allows us to use the boundary conformal field theory technique developed for the multichannel Kondo problem\cite{LA} to obtain an exact description of our newly found intermediate fixed point. The previously mentioned work by Yi and Kane\cite{YK,*Y} also utilized this method to study the intermediate fixed point of the quantum Brownian motion model on the honeycomb lattice. 

Let us recall the Emery-Kivelson solution of the two-channel $SU(2)$ Kondo problem\cite{EK}. This Kondo problem can be mapped to our spinless resonant tunneling problem at $g=1/2$. It was shown that with symmetric channels, at the Toulouse limit, only \textit{half} of the impurity spin degree of freedom is coupled to the conduction electrons resulting in the non-Fermi liquid properties. If we replace the $SU(2)$ impurity spin by the two degenerate charge states of the dot, then the \textit{half} coupling behavior of the Kondo impurity spin at the Toulouse limit is the same as the \textit{half} occupation of the quantum dot by electrons hopping from leads(i.e. $\theta^{\sigma}/\pi$ is a half integer when $K=0$).

Now, for the spinful case, a suitable Kondo problem would be one with two channels and three spin states for the impurity spin. Naturally, this leads us to the two-channel Kondo problem with $SU(3)$ impurity spin for which the three spin states corresponds to the three possible occupation states on the quantum dot(Fig. 9). 

The Hamiltonian of a two-channel $SU(3)$ Kondo problem reads
\begin{equation}
H=iv_F\sum_{s=1}^3 \sum_{a=1}^2\int \psi^{s\dagger}_{a} \partial_x \psi_{a}^{s}+2\pi v_F \sum_{i=1}^8\sum_{a=1}^2 J_i \chi^i S_a^i(x=0), 
\end{equation} where $a$ and $s$ are channel and spin indices respectively and $\vec{\chi}$ is the impurity spin. $SU(3)$ has eight generators $\{\lambda_i\}$, $i=1\dots8$ and thus the electron spin operator $\vec{S_a}=\psi_{as}^{\dagger}(\lambda_{ss'}/2)\psi_{as'}$. We can regroup generators of $SU(3)$ into three pairwise linear combinations of off-diagonal generators in analogy with the $SU(2)$ case as $T_\pm=(\lambda_1 \pm i\lambda_2), U_\pm=(\lambda_4 \pm i\lambda_5)$ and $V_\pm=(\lambda_6 \pm i\lambda_7)$. This allows us to write out the spin operators for electrons. 

Following the Emery-Kivelson solution\cite{EK}, we first bosonize fermions as
\begin{equation}
\psi_{a}^s=\frac{1}{\sqrt{2 \pi v_F \tau_c}}e^{-i\Phi_{a}^s},
\end{equation}
where $\Phi_{a}^s$ is a bosonic field satisfying 
\begin{equation}
[\Phi_{a}^s,\Phi_{a'}^{s'}]=-i \pi \delta_{aa'} \delta^{ss'} \text{sgn} (x-x').
\end{equation} 
Then the operators at each channel are
\begin{eqnarray}
S_a^{T_{\pm}}=\frac{1}{2 \pi v_F \tau_c} e^{\pm i \Phi_a^{\sigma_1}}\\
S_a^{U_{\pm}}=\frac{1}{2 \pi v_F \tau_c} e^{\pm i\frac{(\sqrt{3} \Phi_a^{\sigma_2}+\Phi_a^{\sigma_1})}{2}}\\
S_a^{V_{\pm}}=\frac{1}{2 \pi v_F \tau_c} e^{\pm i\frac{(\sqrt{3} \Phi_a^{\sigma_2}-\Phi_a^{\sigma_1})}{2}}\\
S_a^3=\frac{1}{4\pi} \partial_x \Phi_a^{\sigma_1}\\
S_a^8=\frac{1}{4\pi} \partial_x \Phi_a^{\sigma_2},
\end{eqnarray}
where $\Phi_a^{\sigma_1}=\Phi_a^1-\Phi_a^2$ and $\Phi_a^{\sigma_2}=(1/ \sqrt{3})(\Phi_a^1+\Phi_a^2-2\Phi_a^3)$. If we further assume that our Kondo problem is anisotropic meaning that the diagonal coupling constants $J_3$ and $J_8$ (in analogy with $J_z$ in the $SU(2)$ case) are not equal to the off-diagonal ones for which we call $J_{\perp}$ (in analogy with $J_{\pm}$ in the $SU(2)$ case) in general, then the Hamiltonian becomes
\begin{equation}
H=H_K+H_J
\end{equation}
with 
\begin{equation}
H_K=\sum_{a=1}^2 \frac{v_F}{8 \pi}  \int dx [(\partial_x \Phi_a^{\rho})^2+(\partial_x \Phi_a^{\sigma_1})^2+(\partial_x \Phi_a^{\sigma_2})^2]
\end{equation}
\begin{equation}
\begin{aligned}
H_J&=\frac{1}{2} \sum_{a=1}^2 \left\{v_F\left[J_3 \tau^3 \partial_x \Phi_a^{\sigma_1}(0)+J_8 \tau^8\partial_x \Phi_a^{\sigma_2}(0)\right]\right\}\\
 &+\frac{J_{\perp}}{\tau_c}\sum_{i=T,U,V}[\chi^{i_+} S^{i_-}_a+\text{h.c.}]
\end{aligned}
\end{equation} where $H_K$ is the kinetic energy of electrons and $H_J$ is the interaction between the impurity and electron spins at origin.

Now we introduce a unitary transformation 
\begin{equation}
U_{\epsilon_3, \epsilon_8}=e^{i(\epsilon_3 \sum_a \Phi_a^{\sigma_1}(0)+\epsilon_8 \sum_a \Phi_a^{\sigma_2}(0))},
\end{equation}
to decouple $\partial_x \Phi_a^{\sigma_1}(0)$ and $\partial \Phi_a^{\sigma_2}(0)$ in $H_J$ in Eq. (50) by setting $\epsilon_{3,8}=J_{3,8}/2$.

Then we perform an orthogonal transformation for variables
\begin{equation}
\begin{bmatrix}
\Phi_{sf}^{\sigma_i} \\
\Phi_{s}^{\sigma_i}
\end{bmatrix}
=\textbf{O}\begin{bmatrix}
\Phi_1^{\sigma_i} \\
\Phi_2^{\sigma_i}
\end{bmatrix},
\end{equation} with
\begin{equation}
\textbf{O}= 
\begin{pmatrix}
\frac{1}{\sqrt{2}} & -\frac{1}{\sqrt{2}} \\
 \frac{1}{\sqrt{2}} &\frac{1}{\sqrt{2}}
\end{pmatrix}.
\end{equation}

The partition function of our anisotropic Kondo problem is
\begin{equation}
\begin{aligned}
&Z=\sum_n \frac{1}{n!} (\frac{J_{\perp}}{2})^n \sum_{a=1}^2\int \frac{d^n\tau_j}{\tau_c} \langle(\chi^{T_+}e^{-i(\frac{1-J_3}{\sqrt{2}}\Phi_s^{\sigma_1}+O^{-1}_{a1} \Phi_{sf}^{\sigma_1})}\\
&+\chi^{U_+}e^{-\frac{i}{2}((\frac{1-J_3}{\sqrt{2}}\Phi_s^{\sigma_1}+\frac{\sqrt{3}(1-J_8)}{\sqrt{2}}\Phi_s^{\sigma_2})+(O^{-1}_{a1}\Phi_{sf}^{\sigma_1}+\sqrt{3}O^{-1}_{a1}\Phi_{sf}^{\sigma_2}))}\\
&+\chi^{V_+}e^{-\frac{i}{2}((-\frac{1-J_3}{\sqrt{2}}\Phi_s^{\sigma_1}+\frac{\sqrt{3}(1-J_8)}{\sqrt{2}}\Phi_s^{\sigma_2})+(-O^{-1}_{a1}\Phi_{sf}^{\sigma_1}+\sqrt{3}O^{-1}_{a1}\Phi_{sf}^{\sigma_2}))}\\
&+\text{h.c.})^n\rangle\\
&=\sum_n \frac{1}{n!} (\frac{J_{\perp}}{2\tau_c})^n \sum_{a_i=\pm} \int {d^n \tau_j} e^{\sum_{k<l}V_{kl}}\delta(\sum_p O_{a_p 1}\vec{r_p}),
\end{aligned}
\end{equation}
the interaction potential is given as
\begin{equation}
V_{kl}=2(\vec{r_{k}}\textbf{R}\vec{r_{l}}+O_{a_k 1}^{-1}O_{1 a_l} \vec{r_k} \cdot \vec{r_l}) \ln{\frac{\tau_k-\tau_l}{\tau_c}},
\end{equation} 
\begin{eqnarray}
\vec{r_{T_\pm}}=\pm(1,0)\\
\vec{r_{U_\pm}}=\pm(\frac{1}{2},\frac{\sqrt{3}}{2})\\
\vec{r_{V_\pm}}=\pm(-\frac{1}{2},\frac{\sqrt{3}}{2})
\end{eqnarray} and \textbf{R}=
$\begin{pmatrix}
R_3^2 & 0\\
0 & R_8^2
\end{pmatrix} $ =
$\begin{pmatrix}
\frac{(1-J_3)^2}{2} & 0\\
0 & \frac{(1-J_8)^2}{2}
\end{pmatrix} $. 

\begin{figure}[htbp]
\centering 
 \includegraphics[width=3in]{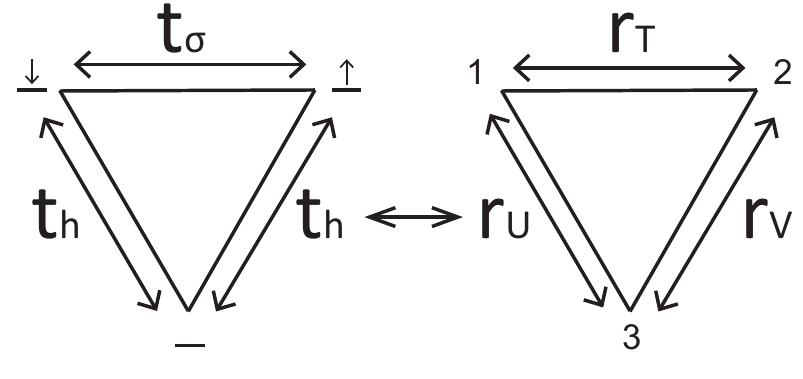}
\caption{Mapping between resonant tunneling problem and Kondo problem. Number 1-3 indicates the three spin states for the impurity spin and \textbf{r}s are spin transferring processes of the Kondo problem.}
\end{figure}

The following mapping turns our resonant tunneling problem to the two-channel $SU(3)$ Kondo problem:
\begin{eqnarray}
& \frac{J_\perp}{2} \leftrightarrow t\\
& {\chi^{i\pm}}_{i=U,V,T} \leftrightarrow \ket{l}\bra{m}_{l\neq m, l,m=\uparrow,\downarrow,0}\\
& (\Phi_{sf}^{\sigma_1},\Phi_{sf}^{\sigma_2},\Phi_{s}^{\sigma_1},\Phi_{s}^{\sigma_2})\leftrightarrow (\varphi^{\rho}_{\sigma},\varphi^{\rho}_{\rho},\varphi^{\sigma}_{\sigma}, \varphi^{\sigma}_{\rho})\\
& \textbf{R} \leftrightarrow \textbf{K}\\
& \vec{r_i} \leftrightarrow \vec{H_i}\\
& O^{-1}_{a_i 1} \leftrightarrow a_i.
\end{eqnarray} The essential idea is to relate the spin and charge transferred in the resonant tunneling problem to spin transferred in the Kondo problem. In this way, when $g_{\rho}=1/3$ and $g_{\rho}=1$, the spin $SU(2)$ symmetry $\times$ $U(1)$ charge symmetry can be mapped to the $SU(3)$ symmetry of the Kondo problem. Then, we can apply the same analysis and conclude that both problems flow to the same intermediate fixed point described above. So if we start off with all $J_{\perp}$s equal and $J_3=J_8$, then the system flows into an intermediate fixed point lying on the Toulouse limit line (Fig. 7) with the full $SU(3)$ symmetry ($J=J_3=J_8=1$) and ($\Phi_s^{\sigma_1}, \Phi_s^{\sigma_2}$) fields decouple.

\section{UNIVERSAL RESONANCE}
In this section, we utilize the previously established mapping to the multichannel Kondo problem to study our tunneling problem on resonance.  As a known result\cite{KF}, at low but finite temperature, the width of the resonance line-shape vanishes as a power of temperature. 

\begin{figure}[htbp]
\centering 
 \includegraphics[width=3.4in]{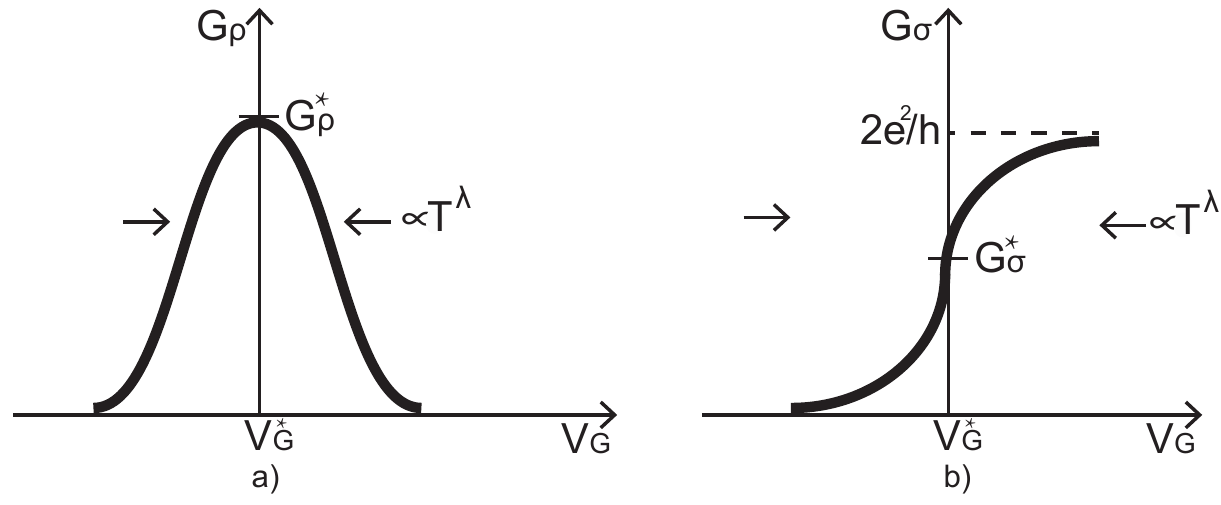}
\caption{Universal scaling function for a) charge conductance and b) spin conductance.}
\end{figure}

The charge and spin conductance through resonance assumes a universal shape as a scaling function\cite{TK} (Fig. 10)
\begin{eqnarray}
G_\rho(\delta,T)=\mathcal{G}_\rho(c \frac{\delta}{T^{\lambda}})\\
G_\sigma(\delta,T)=\mathcal{G}_\sigma(c \frac{\delta}{T^{\lambda}})
\end{eqnarray}
 where $c$ is a non-universal constant and $\delta$ is the distance to resonance in gate voltage. Using boundary conformal field theory description of the Kondo problem, we calculate the on-resonance conductance $G^*_{\rho,\sigma}$ at $T=0$ in Subsection A. Moreover, in Subsection B, we exam all allowed operators at the fixed point of our resonant tunneling problem and identified the one relevant operator for which we could tune the system through resonance. The scaling behavior of the resonance line-shape, which depends on the critical exponent $\lambda=1-\Delta$, is obtained from calculating the scaling dimension $\Delta$ of that relevant operator. 

\subsection{ON-RESONANCE CONDUCTANCE}
\begin{figure}[htbp]
\centering 
 \includegraphics[width=3in]{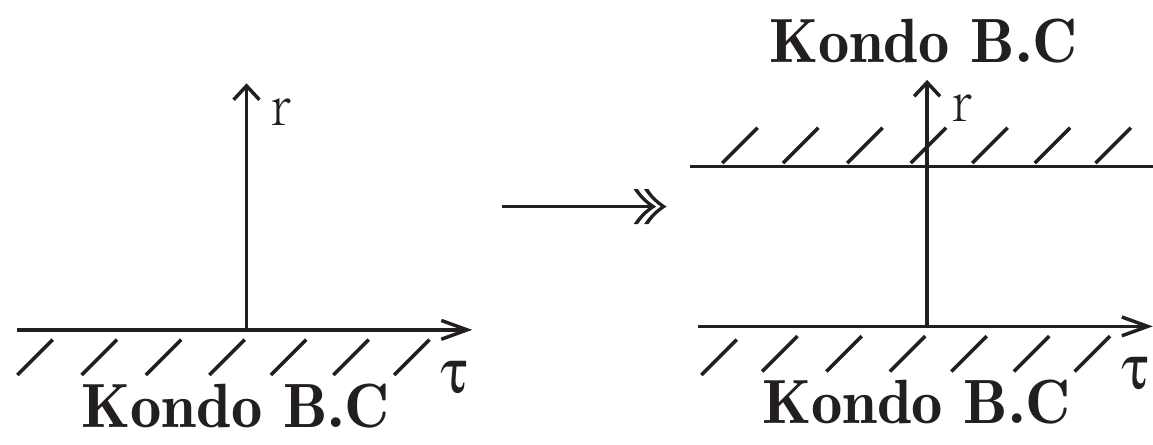}
\caption{Boundary conformal field theory description of Kondo problem}
\end{figure}

Unlike the spinless case in which the fixed point is perturbatively accessible at the small-barrier limit, here, we can not extract information about conductance without an exact description of the intermediate fixed point. Despite the failure of perturbative methods, since our resonant tunneling problem is equivalent to a two-channel $SU(3)$ Kondo problem, we can study the intermediate fixed point using boundary conformal field theory. 

The BCFT description of our Kondo problem resides on the upper half-plane (Fig. 11) with Kondo boundary condition encoded on the real axis\cite{L,LA}. It is precisely this non-trivial Kondo boundary condition that gives calculations of correlation functions an extra twist as reflected in Appendix B. Following Eqs. (56)-(61), by identifying electron operators in our tunneling problem with the two-channel $SU(3)$ Kondo problem, we have our current-operator correspondence:
\begin{eqnarray}
&J^\rho_\sigma(0^\pm,\tau) \leftrightarrow J_{sf}^{\sigma_1}(0^\pm,\tau)\\
&J^\rho_\rho(0^\pm,\tau) \leftrightarrow J_{sf}^{\sigma_2}(0^\pm,\tau).
\end{eqnarray}
Using the Kubo formula, the on-resonance Kubo conductance is 
\begin{eqnarray}
G^{K*}_{\rho}=2g_{\rho}  \mu^* \frac{e^2}{h}\\
G^{K*}_{\sigma}=2g_{\sigma}  \mu^* \frac{e^2}{h}.
\end{eqnarray}
For $g_\rho=1/3$ and $g_\sigma=1$, the mobility $u^*$ is calculated using the boundary conformal field theory in Appendix B. We have  
\begin{equation}
\mu^*=\frac{5-\sqrt{5}}{4}\approx 0.691.
\end{equation}

The physical conductance and its relation to the Kubo conductance calculated above is explained in Appendix C. The physical on-resonance conductance is 
\begin{equation}
G^*_{\rho,\sigma}=\frac{2g_{\rho,\sigma} \mu^*}{1+(g_{\rho,\sigma}-1) \mu^*}\frac{e^2}{h}.
\end{equation}
For $g_\rho=1/3$ and $g_\sigma=1$, we have 

\begin{eqnarray}
G^*_{\rho}\approx 0.854 \frac{e^2}{h}\\
G^*_{\sigma} \approx 1.382\frac{e^2}{h}.
\end{eqnarray}
\subsection{TUNING THROUGH RESONANCE}
According to Cardy\cite{C}, properties of the boundary operators can be obtained by conformally mapping the upper half-plane to an infinite stripe with Kondo boundary conditions on both ends (Fig. 11). Therefore, to obtain the spectrum of Hamiltonian $H_{KK}$ in an infinite stripe with the ``Kondo-Kondo" boundary condition, we can use the ``double fusion" rule hypothesized by Affleck and Ludwig\cite{L,LA} starting with the free fermion boundary condition ``$FF$" on both ends. Since the conformal embedding of our Kondo problem is $U(1)^{charge}\times SU(2)_3^{flavor} \times SU(3)_2^{spin}$, just like the case for the three-channel $SU(2)$ Kondo problem with spin and flavor interchanged, then any boundary operators can be represented as a triplet $(Q,j,\lambda)$ where the three quantum numbers are weights in representations of Lie groups $U(1), SU(2)$ and $SU(3)$ respectively\cite{L,LA}. The allowed triplets are of course all possible primary fields at the intermediate fixed point. The calculation from Eq. (B12) shows that both the two-channel $SU(3)$ and the three-channel $SU(2)$ Kondo problems\cite{YK,*Y} flow to the same intermediate fixed point. Therefore, using the latter Kondo problem, if we start with the free fermion boundary condition $(0,0,0)$, and fuse the boundary operators with the impurity spin operator $s=1/2$ $(0,1/2,0)$ twice, the resultant operators are all possible primary fields at the intermediate fixed point. Their scaling dimensions are given as\cite{DMS} 
\begin{equation}
\Delta=\frac{(\lambda, \lambda+2\rho)}{2(k+g)},
\end{equation}
where $( \cdot )$ is the scalar product induced by Killing forms, $\lambda$ is the weight of the boundary operator in the corresponding representation of its Lie group, $\rho$ is the Weyl vector, $k$ is the level and $g$ is the Coexter number\cite{DMS}. The only relevant operators left are $(0,1,0)$ with $\Delta=1/2$ and $(0,0,[1,1])$ with $\Delta=3/5$ which transform as elements of the adjoint representation of $SU(2)$ and $SU(3)$, respectively. 

Counting the number of available operators is the same as counting the dimension of the two adjoint representations, which gives dim(ad $SU(2)$)+dim(ad $SU(3)$)=11 possible relevant operators. However, channel symmetry and $SU(3)$ spin symmetry in the Kondo problem all impose constraints via conservation laws. 
 Any off-diagonal elements of the adjoint representation will modify either channel number or spin numbers and thus break the conservation laws. We are left with three diagonal relevant operators. 

In the familiar two-channel $SU(2)$ Kondo problem, there are two relevant diagonal operators, each from $SU(2)_2^{spin}$ and $SU(2)_2^{flavor}$ sectors, respectively. Inversion symmetry demands there should be no difference between two channels. Therefore, the diagonal operator from $flavor$ $SU(2)$ can not be present since it will lead the system flow towards an anisotropic Kondo fixed point with one channel strongly coupled and the other disconnected by breaking the $flavor$ $SU(2)$ symmetry\cite{ALPC}. When interpreting this in the spinless resonant tunneling problem, since the fixed point is at the perfect conducting limit, then $\cos(2\theta)$ and $\sin(2\theta)$ are the two aforementioned relevant diagonal operators. Inversion symmetry in this case requires the two barriers to be the same and eliminates $\sin(2\theta)$. Similarly, in the two-channel $SU(3)$ Kondo problem, inversion symmetry again eliminates any relevant diagonal operator from $flavor$ $SU(2)$. Moreover, we know that the relevant diagonal operator in subgroup $spin$ $SU(2) \subseteq spin$ $SU(3)$ must vanish to make $r_U=r_V$ because when translating back to our resonant tunneling problem, time-reversal symmetry requires there be no difference between spin states so that the two $t_h$ processes are equal. Note that $t_{\sigma}$ might have a different amplitude. This extra degree of freedom is precisely controlled by the remaining one relevant diagonal operator from $spin$ $U(1) \subseteq spin$ $SU(3)$ and can be used to tune the system to resonance.

With this knowledge, at finite temperature, we are able to calculate the critical exponent 
\begin{equation}
\lambda=1-\Delta=2/5
\end{equation}
for the resonant line-shape. The exact form of the scaling function $\mathcal{G}$ can be obtained from the Monte-Carol simulation\cite{MKGF}.

\section{LEVEL-RANK DUALITY IN THE QUANTUM BROWNIAN MOTION MODEL}

\begin{figure}[!htbp]
\centering 
 \includegraphics[width=3in]{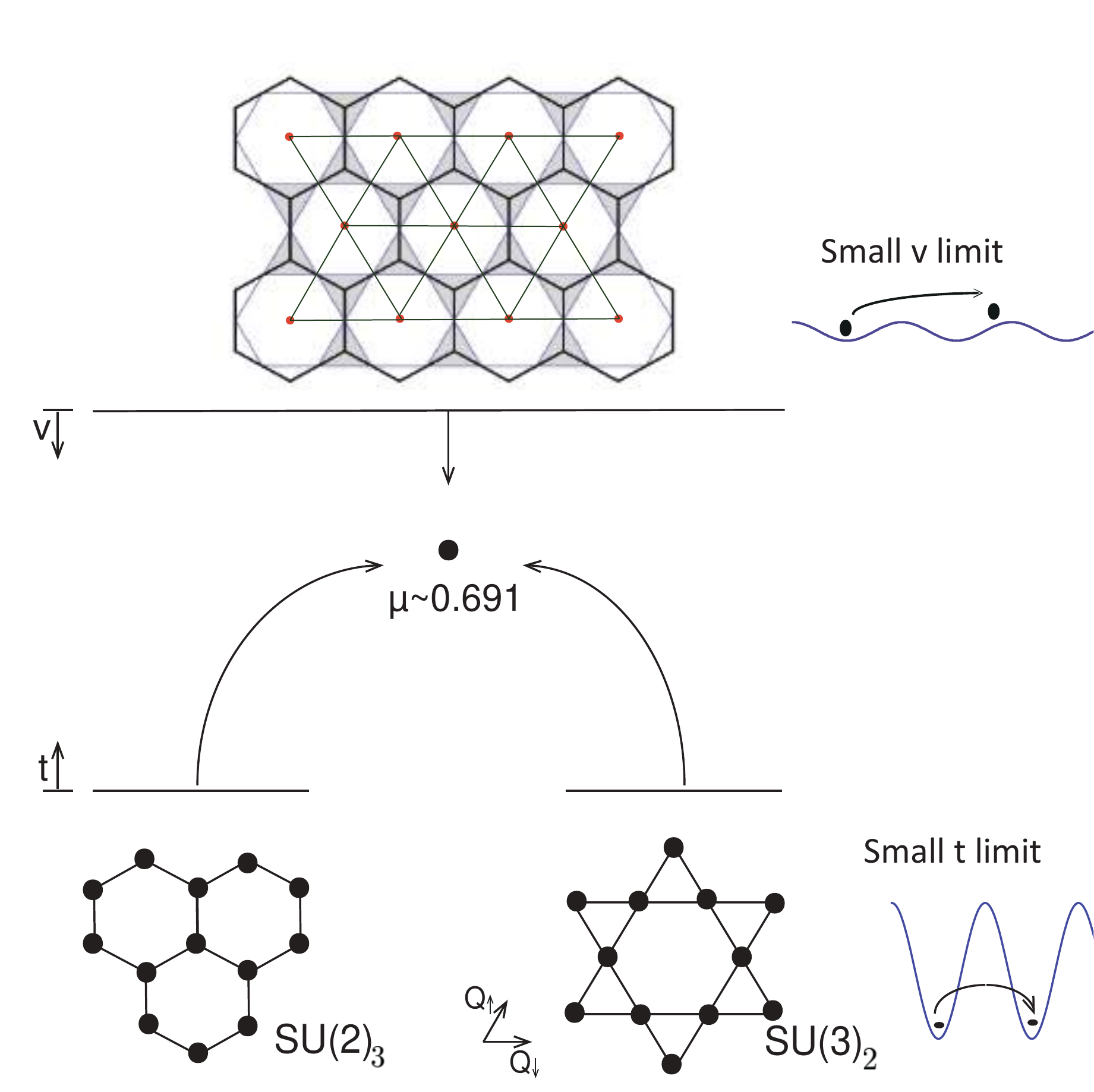}
\caption{Flow diagram for quantum Brownian motion models on a honeycomb and a Kagome lattice. The top(bottom) line represents small(large) barrier limits. Arrows indicate the direction of RG flows and the solid dot represents the intermediate fixed point with its mobility $\mu$ labeled. Since the two models flow to the same intermediate fixed point from both limits, this phenomenon manifests the level-rank duality from a quantum Brownian motion model perspective.}
\end{figure}
BCFT has granted us an exact description of our two-channel $SU(3)$ Kondo fixed point. Translating everything into the quantum Brownian motion on a Kagome lattice using the mapping from Section \Rmnum{3}, the spin-current conductance in the Kondo problem characterizing the intermediate fixed point, becomes the mobility $\mu$ of fictitious particles on the lattice of a quantum Brownian motion model with $\mu=1$ for free particles and $\mu=0$ for completely localized particles (see Appendix B).

The mobility calculated in Eq. (B12) confirms that the quantum Brownian motion on both the honeycomb lattice\cite{YK,*Y} and the Kagome lattice flows into the same strong coupling fixed point (Fig. 12). Mathematically, this can be attributed to the fact that the same conformal embedding is realized at the fixed point, namely $U(1)\times SU(2)_3 \times SU(3)_2$. If we dig in a little further, this phenomenon is called level-rank duality relating $SU(3)_2$ conformal field theory to $SU(2)_3$ conformal field theory \cite{DMS}. However, instead of going through mind-boggling mathematical formalism, here we provide a more physical picture of this equivalency using quantum Brownian motion models. 

This equivalency in the large-barrier limit can be assessed by comparing the mobility of the two quantum Brownian motion models since both are related to Kondo problems in the Toulouse limit. From that calculation, a more general pattern emerges. We find that with $n+k$ fixed, any $SU(n)_k$ quantum Brownian motion model flows into an intermediate fixed point with the same mobility $\mu=2 \sin^2\left[\pi/(n+k)\right]$. This is checked up to $n+k=10$ using MATHEMATICA. 

On the other hand, in the small-barrier limit, to the first order, the quantum Brownian motion model is governed by the renormalization group Eq. (34) which drives the system towards the intermediate fixed point. The general pattern stated in the previous paragraph is harder to establish since for general $n$ and $k$, $v_{\bold{G}}$ are complex numbers ($v_{\vec{G}}=v^*_{-\vec{G}}$). Therefore, it is not clear how the two systems will behave under the first order flow equation. However, for special cases with either $n$ or $k=2$, there is a simple argument to show their equivalency at the small-barrier limit. First of all, it is not hard to see that the $SU(2)_k$ quantum Brownian motion lives on a generalized honeycomb lattice and the $SU(k)_2$ quantum Brownian motion lives on a generalized Kagome lattice both in $k-1$ dimensional space. In Appendix D, by choosing an appropriate origin, we have shown that all $v_{\vec{G}}\in \mathbb{R}$ and $v^h_{\vec{G}_0}$ of the honeycomb lattice have the same sign as $v^k_{\vec{G}_0}$ of the Kagome lattice. Therefore, according to the flow equation, there is no difference between the two quantum Brownian motion models.

Higher dimensional objects are always difficult to envision, so we leave detailed calculations in Appendix D. Here, we will stick to the simplest case with honeycomb and Kagome lattices. A hand-waving explanation would be that if you blur your eyes, there is not much difference between a honeycomb lattice and a Kagome lattice. Of course, we can show this more rigorously. It is known that both honeycomb and Kagome lattices have the triangular lattice as their Bravais lattice. As a result, they also share the same reciprocal lattice. Next, we can ``gauge" $v_{\vec{G}}$ so that $v_{\vec{G}_0}$ is the same for all shortest reciprocal lattice vectors $\vec{G}_0$. This is achieved by shifting the origin of our coordinate system to the center of hexagon (a center of inversion) since $v_{\vec{G}}$ are coordinate dependent. We have $v_{\vec{G}_0}=-1$ for both honeycomb and Kagome lattices. Thus, the quantum Brownian motion on the two lattices behaves the same whenever $\vec{G}$ is a relevant perturbation.

\section{Conclusion}
In this paper, we have studied the problem of resonant tunneling in a spinful Luttinger liquid. We showed that along with the spinless resonant tunneling problem, they both possess symmetry protected topological phases separated by quantum critical points. For the spinless case, two insulating symmetry protected topological phases protected by inversion symmetry are found and characterized by a polarization defined as the number of charge transferred across the infinite barrier. Similarly, for $1/2<g_\rho<1$ and $g_{\sigma}=1$, 
two charge and spin insulating symmetry protected topological phases protected by inversion and time-reversal symmetry are found and characterized by a polarization defined as the number of pairs of electrons with opposite spin transferred across the infinite barrier. By tuning the system through perfect resonance, we can go through quantum critical points and access all phases. The most interesting phase emerges when $g_\rho<1/2$. It is a charge insulating but spin conducting phase. Its transitions to other insulating phases are governed by an intermediate fixed point exhibiting non-Fermi liquid properties.

In order to study this intermediate fixed point, we adjust Luttinger parameters to $g_\rho=1/3$ and $g_\sigma=1$. We solved RG flow equations for our resonant tunneling problem at this setting. At the Toulouse limit, we mapped it to a quantum Brownian motion model on a Kagome lattice and identified the stable intermediate fixed point. With the insight from Yi and Kane\cite{YK,*Y}, we found that the quantum Brownian motion model is precisely the Toulouse limit of a two-channel Kondo problem with $SU(3)$ impurity spin. With the equivalency between the Kondo problem and our resonant tunneling problem established, this allowed us to take advantage of the boundary conformal field theory solution of the multichannel Kondo problem. We obtained an exact description of our intermediate fixed point and calculated the on-resonance conductance as well as the scaling behavior of the resonance line-shape as a function of temperature. 

Finally, we discussed the equivalence between the two-channel $SU(3)$ Kondo problem($SU(3)_2$ CFT) and the three-channel $SU(2)$ Kondo problem ($SU(2)_3$ CFT) from a quantum Brownian motion perspective. We also made a generalization establishing the equivalence between $SU(k)_2$ and $SU(2)_k$ Kondo problems.

\begin{acknowledgments}

\end{acknowledgments}
We thank A. W. W. Ludwig, A. Stern, P. Fendley, S. H. Simon for helpful discussion and comments. This work was supported by a Simons Investigator grant to C.L.K. by the Simons Foundation.

\appendix

\section{Renormalization group for resonant tunneling problem in a spinful luttinger liquid}
Here we set $g_{\rho}=1/3$ and $g_{\sigma}=1$ and adopt the renormalization group calculation developed by Anderson, Yuval, and Hanmann\cite{AYH}. 

First, we decimate all possible closely placed pairs of charges with a range between $\tau_c$ and $\tau_c+d\tau_c$. Inserting them in between charge $i$ and $i+1$, the partition function becomes
\begin{equation}
\begin{split}
&Z=\sum_{n} \frac{1}{n!}t^n \int {d^n \tau_i} \sum_{a_i} e^{-\sum_{i<j}V_{ij}} \\ 
&\times [1-t^2 d\tau_c \sum_{i} \int_{\tau_i+\tau_c}^{\tau_{i+1}+\tau_c} d \tau \sum_{a} e^{V_{ia}(\tau)}+\cdots]
\end{split}
\end{equation} where the interaction of the dipoles with all other charges is
\begin{equation}
V_{ia}(\tau)=-\sum_j \sum_{\vec{r}}[\vec{H} \textbf{K}^2 \vec{H_j}+a a_j \vec{H}\cdot \vec{H_j}] \tau_c \partial_\tau \ln{\frac{\tau-\tau_j}{\tau_c}}.
\end{equation} Expand the exponent
\begin{equation}
\begin{split}
&Z=\sum_{n}\frac{1}{n!} t^n \int {d^n \tau_i} \sum_{a_i} e^{\sum_{i<j}V_{ij}}\\
&\times [1-t^2 d\tau_c \sum_{i} \int_{\tau_i}^{\tau_{i+1}} d \tau \sum_{a} (1+V_{ia}(\tau)+\cdots)+\cdots]. 
\end{split}
\end{equation}
\begin{figure}[htbp]
\centering 
 \includegraphics[width=1in]{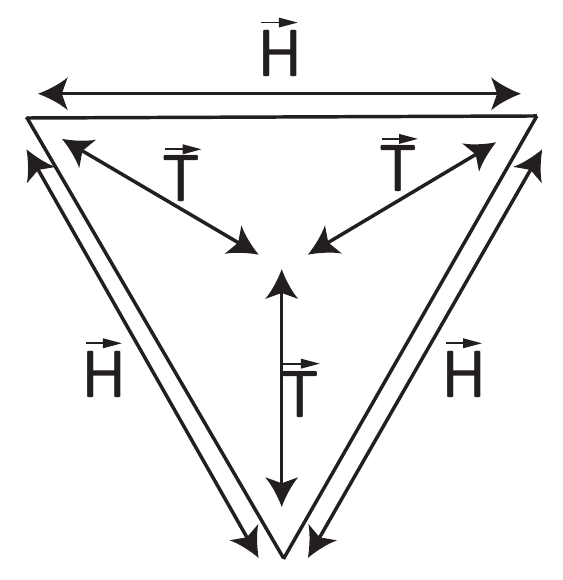}
\caption{Hopping vectors $\vec{H}$ and auxiliary vectors $\vec{T}$ }
\end{figure}

Now, define $\vec{H_i}=\vec{T}_{i+1/2}-\vec{T}_{i-1/2}$ (Their relation is depicted in Fig. 13), 
\begin{equation}
\begin{split}
&\sum_{i,a}\int_{\tau_i}^{\tau_{i+1}} V_{ia}(\tau)\\
&=-\sum_{ij}\sum_{\vec{H}}2\vec{H} \textbf{K} \vec{H_j} \tau_c (\ln{\frac{\tau_{i+1}-\tau_j}{\tau_c}}-\ln{\frac{\tau_i-\tau_j}{\tau_c}})\\
&=-\sum_{i<j} \sum_{\vec{T}\neq \vec{T}_{i+\frac{1}{2}}}4(\vec{T}-\vec{T}_{i+\frac{1}{2}}) \textbf{K} \vec{H_j} \tau_c (\ln{\frac{\tau_{i+1}-\tau_j}{\tau_c}}\\
&-\ln{\frac{\tau_i-\tau_j}{\tau_c}})\\
&=-\sum_{i<j} \sum_{\vec{T}}4(\vec{T}-\vec{T}_{i+\frac{1}{2}}) \textbf{K} \vec{H_j} \tau_c (\ln{\frac{\tau_{i+1}-\tau_j}{\tau_c}}\\
&-\ln{\frac{\tau_i-\tau_j}{\tau_c}})\\
&=-\sum_{i<j} (\sum_{\vec{T}}4\vec{T}-3\cdot4\vec{T}_{i+\frac{1}{2}}) \textbf{K} \vec{H_j} \tau_c (\ln{\frac{\tau_{i+1}-\tau_j}{\tau_c}}\\
&-\ln{\frac{\tau_i-\tau_j}{\tau_c}})\\
&=-\sum_{i<j} 12\vec{H_i} \textbf{K} \vec{H_j} \tau_c \ln{\frac{\tau_{i}-\tau_j}{\tau_c}}
\end{split}
\end{equation}

Due to neutrality condition, $\sum_i \vec{H_i}=0$ gives $\sum_{i<j} \vec{H_i} \cdot \vec{H_j}=-1/2n$. Rescaling $\tau_c$ to $\tau_c+d\tau_c$, the partition function becomes
\begin{equation}
\begin{aligned}
&Z=\sum_n \sum_{\{a_i\}} t^n (\frac{\tau_c+d\tau_c}{\tau_c})^{-\frac{1}{2}n(1+K^2)}\\
&\times \int d^n\tau_i \text{exp} (-\sum_{i<j} [K^2 \vec{H_i} \cdot \vec{H_j}(1-12t^2 \tau_c d\tau_c)\\
&+a_i a_j \vec{H_i}\cdot \vec{H_j}] \ln{\frac{\tau_i-\tau_j}{\tau_c}}),
\end{aligned}
\end{equation}
which leads to the flow equations
\begin{eqnarray}
&\frac{dt}{d\ell}=(1-\frac{1}{2}(K^2+1))t\\
&\frac{dK}{d\ell}=-6 \tau_c^2 t^2K
\end{eqnarray}
\section{Mobility of quantum Brownian motion on Kagome Lattice}
In this section, we calculate the universal mobility $\mu^*$ of quantum Brownian motion model on Kagome lattice at the intermediate fixed point using boundary conformal field theory results on Kondo problem. The analog of $R_a^i$ is the spin in our Kondo problem $S_a^i$. Therefore spin currents for our Kondo problem are
\begin{equation}
J_a^{\sigma_i}=\frac{v_F}{4\pi}\partial_x \Phi_a^{\sigma_i}=\frac{1}{4\pi}\partial_t \Phi_a^{\sigma_i}
\end{equation} and their linear combinations are
\begin{eqnarray}
&J_s^{\sigma_i}=\sum_{a=1}^2 \frac{1}{\sqrt{2}}J_a^{\sigma_i}=\frac{1}{4\pi}\partial_t\Phi_s^{\sigma_i}\\
&J_{sf}^{\sigma_i}=\sum_{a=1}^2O_{1a}J_a^{\sigma_i}=\frac{1}{4\pi}\partial_t\Phi_{sf}^{\sigma_i}
\end{eqnarray}
can be translated into velocity of Brownian particles\cite{YK,*Y}.

For example, the velocity $\partial_t R_{\sigma_i}$ is mapped to the rate of spin $\sigma_i$ injected into spin-flavor channel
\begin{equation}
J_{sf}^{\sigma_i}(x=0^+)- J_{sf}^{\sigma_i}(x=0^-) \leftrightarrow \partial_t R_{\sigma_i}
\end{equation}
where \begin{equation}
R_{\sigma_i}=\int dx \psi^{\dagger}_{as}\frac{\lambda_{ss'}^{\sigma_i}}{2}\frac{\sigma^z_{aa'}}{2} \psi_{a's'}
\end{equation}

The mobility $\mu$ now becomes the response of spin currents to applied potentials.
\begin{equation}
\begin{aligned}
\mu^*&=\lim_{\omega \rightarrow 0} \frac{1}{2 \pi |\omega|} \int d\tau (1-e^{i\omega \tau})\langle T_\tau[J_{sf}^{\sigma_i}(0^+,\tau)\\
&- J_{sf}^{\sigma_i}(0^-,\tau)][J_{sf}^{\sigma_i}(0^+,0)- J_{sf}^{\sigma_i}(0^-,0)] \rangle_0
\end{aligned}
\end{equation}
where $\langle \cdots \rangle_0$ is the average with respect to the free non-interacting Hamiltonian. 

\begin{figure}[htbp]
\centering 
 \includegraphics[width=3in]{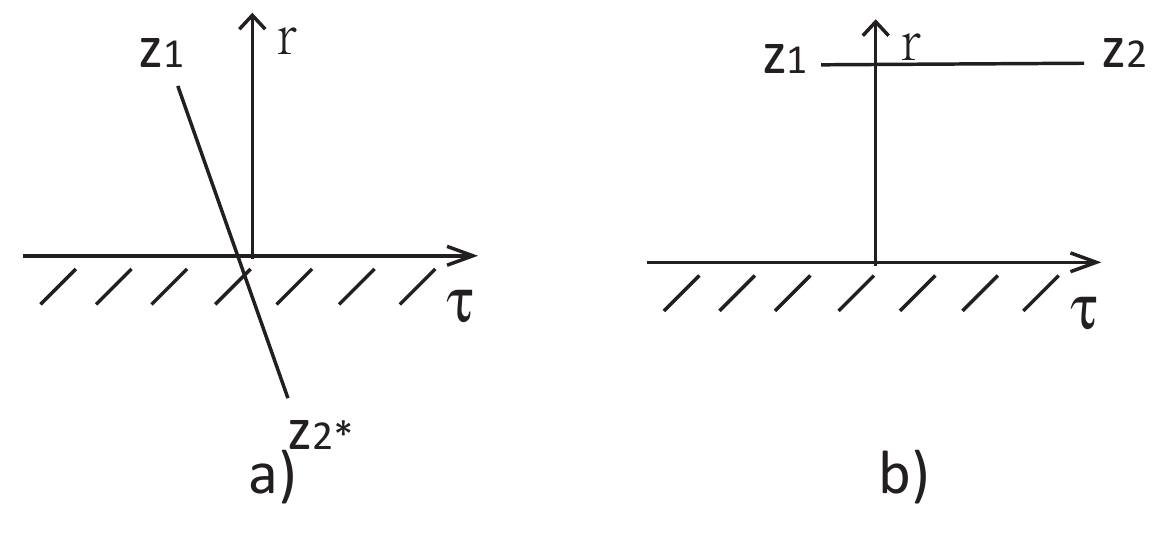}
\caption{a) Correlators like $\langle T_\tau J_{sf}^{\sigma_i}(0^+,\tau)J_{sf}^{\sigma_i}(0^-,0)\rangle_0$ that across the non-trivial boundary can not be translated asymptotically away from the boundary. While in b), correlators like $\langle T_\tau J_{sf}^{\sigma_i}(0^+,\tau)J_{sf}^{\sigma_i}(0^+,0)\rangle_0$ can and thus produce trivial value which is not affected by the boundary.}
\end{figure}

Using boundary conformal field theory\cite{L,LA}, the correlation functions are calculated
\begin{equation}
\begin{aligned}
&\langle T_\tau J_{sf}^{\sigma_i}(0^+,\tau)J_{sf}^{\sigma_i}(0^+,0)\rangle_0=\langle T_\tau J_{sf}^{\sigma_i}(0^-,\tau)J_{sf}^{\sigma_i}(0^-,0)\rangle_0\\
&=\frac{1}{2\tau^2},
\end{aligned}
\end{equation}
\begin{equation}
\begin{aligned}
&\langle T_\tau J_{sf}^{\sigma_i}(0^+,\tau)J_{sf}^{\sigma_i}(0^-,0)\rangle_0=\langle T_\tau J_{sf}^{\sigma_i}(0^-,\tau)J_{sf}^{\sigma_i}(0^+,0)\rangle_0\\
&=\frac{a}{2\tau^2},
\end{aligned}
\end{equation}
where $a$ is a universal complex number depending on Kondo boundary condition and can be calculated using modular $S$-matrix\cite{L,LA} 
\begin{equation}
a=\frac{S^{\lambda}_s/S^{\lambda}_0}{S^{0}_s/S^{0}_0}.
\end{equation}
($\lambda$ is the highest weight representation of $J_{sf}^{\sigma_i}$ and $s$ is the highest weight of representation of impurity spin in the corresponding Lie algebra ). 
 
The general formula for calculating modular $S$-matrix is given as\cite{DMS} 
\begin{equation}
\begin{aligned}
S^{\lambda}_s&=i^{|\triangle_+|}|P/Q^{\vee}|^{-\frac{1}{2}}(k+g)^{-r/2}\\
&\times \sum_{w \in W} \epsilon(w) e^{-2\pi i(w(\lambda+\rho),s+\rho)/(k+g)}
\end{aligned}
\end{equation}
where $|\triangle_+|$ is the number of positive roots, $|P/Q^{\vee}|=det A_{ij}$($A$ is the Cartan matrix) for simply-laced algebras, $k$ is the level, $g$ is the dual Coexter number, $W$ is the Weyl group, $\epsilon(w)$ is the signature function and $\rho$ is the Weyl vector which is half of the sum of all positive roots. 

For our $SU(3)$ case, we have $J_{sf}^{\sigma_i}$ is in the adjoint representation of $SU(3)$ and the impurity spin is in the fundamental representation. Thus, $\lambda=[1,1]$ and $s=[1,0]$ or $[0,1]$. Then 
\begin{equation}
 a=\frac{(-4\sin{\frac{2\pi}{5}}+2\sin{\frac{4\pi}{5}})/(2\sin{\frac{2\pi}{5}}+4\sin{\frac{4\pi}{5}})}{(4\sin{\frac{4\pi}{5}}+2\sin{\frac{2\pi}{5}})/(4\sin{\frac{2\pi}{5}}-2\sin{\frac{4\pi}{5}})}=\frac{\sqrt{5}-3}{2}
\end{equation} which agrees with $a$ calculated in three-channel $SU(2)$ Kondo problem\cite{YK,*Y}.

The mobility is
\begin{equation}
\begin{aligned}
\mu^*&=\frac{1-a}{2}=\frac{5-\sqrt{5}}{4}\\
&\approx 0.691
\end{aligned}    
\end{equation}
\section{Physical conductance}
It is known that the Kubo conductance computed from linear response theory does not match the physical DC conductance measured in a system with Fermi liquid lead\cite{IHJ,DLM,P, K, COA, *COA2,CF}.    The Kubo conductance describes the response of an infinite Luttinger liquid, where the limit $L\rightarrow\infty$ is taken {\it before} $\omega\rightarrow 0$, and relates the current to the potential difference between the incident chiral modes of the Luttinger liquid.     However, the potential of the chiral modes is not the same as the potential of the Fermi liquid leads.  There is a contact resistance between the Luttinger liquid and the electron reservoir where the voltage is defined.   An appropriate model to account for this is to consider a 1D model for the leads in which the Luttinger parameter $g_\rho=g_\sigma=1$ in the leads\cite{CF}.    Here we review that argument for the simple case of spinless electrons characterized by a single Luttinger parameter $g$.   The generalization to include spin is straightforward.

The relationship between the Kubo conductance and the physical conductance can be determined by specifying the appropriate boundary condition at the interface between the Luttinger liquid and Fermi liquid, where $g = g(x)$ in Eq. (3) changes.   Charge conservation requires $\dot\theta$ is continuous, while the condition of zero backscattering at the interface requires $\dot\varphi$ is continuous.   Using the equations of motion determined by Eqs. 1 and 3, we thus conclude that $g v \partial_x\varphi$ and $v\partial_x\theta/g$ are continuous.   Since the Kubo conductance relates the current to the  potential difference between the incoming {\it chiral} modes, it is useful to rewrite this boundary condition in terms of the chiral potentials $V_{R/L} = v(\partial_x\varphi \pm  \partial_x\theta/g)$.   We thus require the continuity of  $g(x)(V_R - V_L)$ (charge conservation) and $V_R+V_L$ (no backscattering). 

\begin{figure}[htbp]
\centering 
 \includegraphics[width=2.5in]{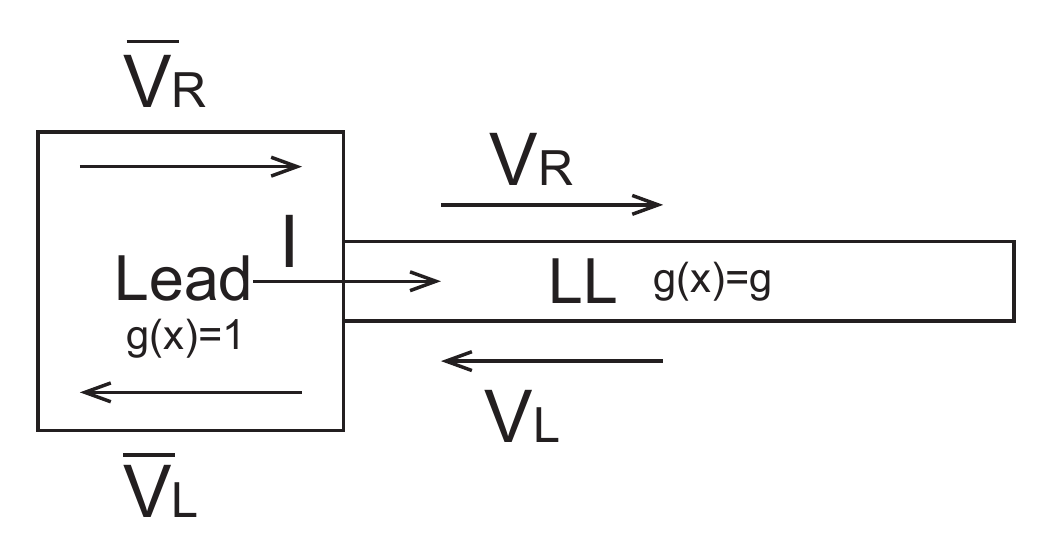}
\caption{Chiral currents in the Fermi-liquid lead and the Luttinger liquid.}
\end{figure}

Applied to a single interface between $g(x)=1$ and $g(x)=g$ (Fig. 15), we thus conclude
\begin{eqnarray}
\bar V_R - \bar V_L &=& g(V_R-V_L) = (h/e^2) I \\
\bar V_R +\bar V_L &=& V_R + V_L.
\end{eqnarray}
Elinminating $\bar V_L$ and $V_L$ leads to
\begin{equation}
\bar V_R - V_R = {h\over e^2} {{g-1}\over{2g}} I
\end{equation}
Thus, the potential of the incoming chiral mode in the Fermi liquid lead is higher than potential of the chiral mode in the Luttinger liquid.   The contact between $g(x)=1$ and $g(x)=g$ effectively contributes a {\it series resistance}
\begin{equation}
R_c = {{g-1}\over{2g}} {h\over e^2}.
\end{equation}

In a two terminal setup with two Fermi liquid leads the series contact resistance is doubled.    Writing the Kubo conductance as $g \mu e^2/h$, where $0<\mu<1$ is the mobility, we then conclude the physical conductance is
\begin{equation}
G = {e^2\over h} { g \mu \over{1 + (g-1)\mu}}.
\end{equation}
This reproduces the fact that for perfect transmission $\mu = 1$ the physical conductance is $e^2/h$, while the Kubo conductance is $g e^2/h$.

For the spinful case, in both the charge and the spin sectors, the contact resistance gets a factor of $1/2$ and the Kubo conductance gets a factor of $2$. Therefore, the physical conductance gets an overall factor of $2$.

\section{Generalized honeycomb and Kagome lattices in small-barrier limit}
In this section we will discuss how our quantum Brownian motion picture of level-rank duality fits to general values of $n$ and $k$ and apply the knowledge to establish proof of equivalence between quantum Brownian motion models on generalized honeycom and Kagome lattice. 

First, notice that we can represent primitive vectors of Bravais lattice of the $SU(n)_k$ quantum Brownian motion model as $n\times k$ matrices $N$ with the only non-zero terms at $N_{11}=N_{ij}=1$ and $N_{1j}=N_{i1}=-1$ for $i,j>1$
\begin{equation}
\frac{1}{2}
\begin{pmatrix}
1 & \dots & -1 & \dots \\
\vdots & \ddots &\vdots & \ddots\\
-1 & \dots & 1 & \dots \\
\vdots & \ddots &\vdots & \ddots
\end{pmatrix} 
\end{equation}
Basically, a primitive vector hops particle between adjacent lattice site with the same basis label. In Kondo language, it refers to processes which leave intact the impurity spin. Since we can choose arbitrary spin state for our impurity spin from $1 \cdots n$, let it be $1$, then each $N$ matrix represents the process which transfers an $i$ spin from channel $1$ to channel $j$ via impurity spin. 

Clearly we can not put $N_{ij}=1$ in either the first row or the first column, it leaves only $(n-1)(k-1)$ independent spots. Therefore, the matrices are actually describing a $(n-1)(k-1)$  dimensional lattice. For the corresponding reciprocal lattice, we find primitive vectors are $n \times k$ matrix $G$:
$$
\pm \frac{2}{nk}
\begin{pmatrix}
1 &  \dots & 1 & -(k-1) \\
 \vdots & \ddots &&\vdots \\
1 &  & \ddots &-(k-1)\\
-(n-1) & \dots & -(n-1) & (n-1)(k-1)
\end{pmatrix}.$$ Different primitive vectors can be obtained by shifting both the row and column containing entry $(n-1)(k-1)$ around at all $nk$ positions in the matrix. 

With the foundation laid, now let us proceed to the special cases with either $n$ or $k=2$. We will show in the following that all $v_{\vec{G}}\in \mathbb{R}$ and $v^{h}_{\vec{G}_0}=c v^{k}_{\vec{G}_0}$ for some positive constant $c$. 

The first trick is to choose the right origin. In general, $v_{\bold{G}}$ will be a complex number, however, if we choose the origin of our coordinate system to be at one of the center of inversion, then we are putting $\vec{G}$ and $-\vec{G}$ at the same footing. Therefore, $v_{\vec{G}}=v_{-\vec{G}}$ which makes it a real number. For later calculation convenience, we will choose one of the center of the bond of our generalized honeycomb lattice to be the origin for both lattices (Fig. 16).

\begin{figure}[htbp]
\centering 
 \includegraphics[width=1.5in]{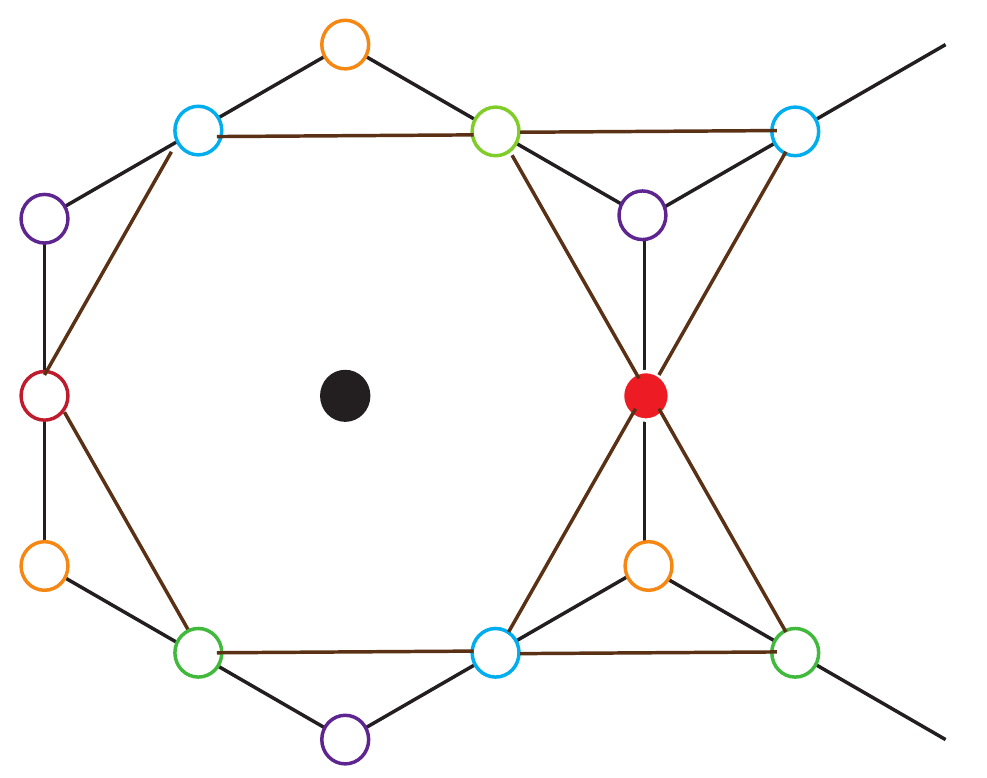}
\caption{Two choices of center of inversion for both lattices: 1. the center of hexagon(black), 2. the center of a bond of the honeycomb lattice(red). However, it is hard to define an analogous position of the first one for generalized honeycomb lattice in odd dimensional space.}
\end{figure}

Next, let us again embed our $k-1$ dimensional lattices into a higher dimensional space, this time a $k$ dimensional space. For generalized $k$-honeycomb and Kagome lattice, we found basis vectors are:
\begin{equation}
\begin{aligned}
&\vec{l}^h_{1}=\frac{1}{2} ( 1,\overbrace{\cdots}^\text{0s}),
&\vec{l}^h_{2}=\frac{1}{2} (- 1,\overbrace{\cdots}^\text{0s})
\end{aligned}
\end{equation}and 
\begin{equation}
\begin{aligned}
&\vec{l}^k_{i=1,\dots,k-1}=\frac{1}{2} (1,\overbrace{\cdots}^\text{0s},\underset{i+1}{-1},\overbrace{\cdots}^\text{0s}),
&\vec{l}^k_{k}=0,
\end{aligned}
\end{equation}
and the shortest vectors for their reciprocal lattice are:
\begin{equation}
\vec{G}_0^{j=1,\dots,k}=\frac{2\pi}{k} (\overbrace{\cdots}^\text{1s},\underset{j}{-k+1},\overbrace{\cdots}^\text{1s}).
\end{equation}

What's left is just plug and chug. Substitute our vectors into 
\begin{equation}
v_{\vec{G}}=\sum_{\vec{G}} e^{i \vec{G} \cdot \vec{l}}
\end{equation}

 we find

\begin{equation}
\begin{aligned}
&v^h_{\vec{G}_0^j}=\begin{cases}
    2\cos{\frac{\pi}{k}},& \text{if } j \neq 1\\
    -2\cos{\frac{\pi}{k}},              & \text{otherwise}
\end{cases}
\end{aligned}
\end{equation}and 
\begin{equation}
\begin{aligned}
&v^k_{\vec{G}_0^j}=\begin{cases}
    k-2,& \text{if } j \neq 1\\
    2-k,              & \text{otherwise}
\end{cases}
\end{aligned}.
\end{equation}

Since $k\in \mathbb{Z}$ and $k \geq 2$, we conclude that $v^h_{\vec{G}_0}$ has the same sign as $v^k_{\vec{G}_0}$.


\begin{thebibliography}{56}
\bibitem [{Note$*$()}]{Note$*$}%
  \BibitemOpen
  \bibinfo {note} {The value of Luttinger parameters $g_{\rho }$ and $g_{\sigma
  }$ is set to $2$ for noninteracting electrons in spinful Luttinger liquid in
  Ref. \protect \rev@citealpnum {KF2,KF}}\BibitemShut {NoStop}%
\end{thebibliography}
\end{document}